\documentclass[1p]{elsarticle}
\journal{Computer Physics Communications}

\newcounter{bla}

\biboptions{comma,square,sort&compress}
\usepackage{hyperref}
\hypersetup{pdfborder={0 0 0},
            colorlinks=true,
            urlcolor=blue,
            breaklinks=true,
            extension = }
\usepackage{physics}
\usepackage{amssymb}
\usepackage{amsmath}
\usepackage{mathtools}
\usepackage{wasysym}
\DeclareMathSymbol{\shortminus}{\mathbin}{AMSa}{"39}

\usepackage[version=4]{mhchem}

\usepackage{algorithm}
\usepackage{algpseudocode}

\makeatletter
\renewcommand\theHALG@line{\thealgorithm.\arabic{ALG@line}}
\makeatother

\makeatletter
\def\ALG@step%
   {%
   \refstepcounter{ALG@line}%
   \stepcounter{ALG@rem}%
   \ifthenelse{\equal{\arabic{ALG@rem}}{\ALG@numberfreq}}%
      {\setcounter{ALG@rem}{0}\alglinenumber{\arabic{ALG@line}}}%
      {}%
   }%
\makeatother

\makeatletter
\newcommand{\ALG@lineautorefname}{L\@gobble}
\makeatother

\usepackage{float}
\usepackage[caption=false]{subfig}

\usepackage{lineno}

\usepackage{booktabs}

\usepackage{xspace}

\newcommand{\Calzone}{\textproc{Calzone}\xspace}
\newcommand{\DeepL}{\textproc{DeepL}\xspace}

\newcommand{\Fluka}{\textproc{Fluka}\xspace}

\newcommand{\Geant}{\textproc{Geant4}\xspace}
\newcommand{\Goupil}{\textproc{Goupil}\xspace}
\newcommand{\MCNP}{\textproc{MCNP}\xspace}
\newcommand{\Penelope}{\textproc{Penelope}\xspace}
\newcommand{\Turtle}{\textproc{Turtle}\xspace}

\newcommand{\C}{C\xspace}

\newcommand{\PyO}{PyO3\xspace}
\newcommand{\Python}{Python\xspace}
\newcommand{\Rust}{Rust\xspace}

\usepackage{acronym}
\newacro{ABI}[ABI]{Application Binary Interface}
\newacro{AMC}[AMC]{Adjoint Monte Carlo}
\newacro{API}[API]{Application Programming Interface}
\newacro{BMC}[BMC]{Backward Monte Carlo}
\newacro{CC}[CC]{Charged Current}
\newacro{CDF}[CDF]{Cumulative Distribution Function}
\newacro{CI}[CI]{Continuous Integration}
\newacro{CM}[CM]{Center of Mass}
\newacro{DCS}[DCS]{Differential Cross-Section}
\newacro{DEM}[DEM]{Digital Elevation Model}
\newacro{IS}[IS]{Importance Sampling}
\newacro{KDE}[KDE]{Kernel Density Estimate}
\newacro{Lips}[Lips]{Lorentz Invariant Phase Space}
\newacro{PDF}[PDF]{Probability Density Function}
\newacro{PDG}[PDG]{Particle Data Group~\cite{Workman2022}}

\usepackage{etoolbox}
\makeatletter
\patchcmd{\ps@pprintTitle}% <cmd>
  {Preprint submitted to}% <search>
  {Accepted for publication in}% <replace>
  {}{}% <succes><failure>
\patchcmd{\ps@pprintTitle}% <cmd>
  {\@date}% <search>
  {April 29, 2025}% <replace>
  {}{}% <succes><failure>
\makeatother

\begin{document}
\begin{frontmatter}
\title{Goupil: A Monte Carlo engine for the backward transport of low-energy
    gamma-rays}

\author[lpc]{Valentin~Niess\corref{cor1}}\ead{niess@in2p3.fr}
\author[lpc]{Kinson~Vernet}
\author[lpc,lmv]{Luca~Terray}

\cortext[cor1]{Corresponding author}

\address[lpc]{
Universit\'e Clermont Auvergne, CNRS, LPCA, F-63000 Clermont-Ferrand, France.}

\address[lmv]{
Universit\'e Clermont Auvergne, CNRS, IRD, OPGC, Laboratoire Magmas et Volcans,
    F-63000 Clermont-Ferrand, France.}

\begin{abstract}
\Goupil is a software library designed for the Monte~Carlo transport of
low-energy gamma-rays, such as those emitted from radioactive isotopes. The
library is distributed as a \Python module. It implements a dedicated backward
sampling algorithm that is highly effective for geometries where the source size
largely exceeds the detector size. When used in conjunction with a conventional
Monte~Carlo engine (i.e., \Geant), the response of a scintillation detector to
gamma-active radio-isotopes scattered over the environment is accurately
simulated (to the nearest percent) while achieving events rates of a few kHz
(with a $\sim$$2.3\,$GHz CPU).

\\

% Computer program description
\noindent \textbf{PROGRAM SUMMARY}

\begin{small}
\noindent
{\em Program Title: Goupil} \\
{\em CPC Library link to program files:} (to be added by Technical Editor) \\
{\em Developer's repository link: https://github.com/niess/goupil } \\
{\em Code Ocean capsule:} (to be added by Technical Editor)\\
{\em Licensing provisions: LGPL-3.0 } \\
{\em Programming language: \C, \Python and \Rust.} \\
{\em Nature of problem:
    Backward Monte~Carlo transport of gamma-rays that are emitted by
    mono-energetic sources distributed in space.} \\
{\em Solution method:
    A simple modification to a previously presented backward Monte~Carlo
    algorithm~[1].
} \\

\end{small}
\end{abstract}

\end{frontmatter}

%% Add linenumbers
%% \linenumbers

\section{Introduction \label{sec:introduction}}

The emission of gamma photons during the decay of radioactive isotopes offers a
direct method of detecting radioactivity in the environment. Gamma photons have
a long range as compared to other particles emitted by radionuclides, such as
alpha particles and electrons. The observation of gamma photons at a specific
location can indicate the presence of radioactivity at distances ranging from a
few tens of centimetres in aquatic environments and soils to several hundred
metres in the air. This property has been extensively studied in recent decades,
making in-situ gamma-ray spectrometry (as opposed to laboratory gamma-ray
spectrometry) one of the most widely used and accepted methods for the rapid
identification and quantification of radionuclides in the environment. Gamma-ray
measurements have numerous applications in the geosciences, such as mapping
U-Th-K levels on continental surfaces (\citet{Grasty1975}), detecting
radioactive contamination on land (\citet{Sanada2015}), the air
(\citet{Grasty1997}) or at sea (\citet{Povinec1996}), measuring radon in the
soil (\citet{Zafrir2011}), the atmosphere (\citet{Baldoncini2017}) or the
hydrosphere (\citet{Dulai2016}), studying the removal of aerosols by rain
(\citet{Takeuchi1982}), and monitoring soil moisture (\citet{Reinhardt2019}).
Thus, depending on the objectives being pursued, gamma-ray measurements are
performed at -or below- ground level, from the atmosphere using a variety of
aircraft, or underwater.

In a recent study, it has been shown that radon emissions accompanying volcanic
outgassing can be very high, which may indicate internal variations in the
volcano's plumbing system (\citet{Terray2020}). Thus, it is important to have
appropriate tools to continuously monitor radon outgassing into the atmosphere
through volcanic gas plumes. Traditional techniques for measuring radon in the
air rely on the detection of alpha particles, requiring a direct contact between
the gas and the detector. However, due to the aggressive nature of the
environment (high acidity and humidity), traditional techniques are unsuitable
in this context. Therefore, as part of an interdisciplinary project combining
volcanology and nuclear physics, we recently developed a ground-based (in-situ)
gamma-ray spectrometer with a field of view covering $\sim$$2\pi$ solid angle.
Its purpose is to record the flux of descending gamma photons in an attempt to
continuously monitor the activity of radon in the air volume above the volcano.
A detailed description of this spectrometer will be given in a later article.

\begin{figure}[th]
    \center
    \includegraphics[width=\textwidth]{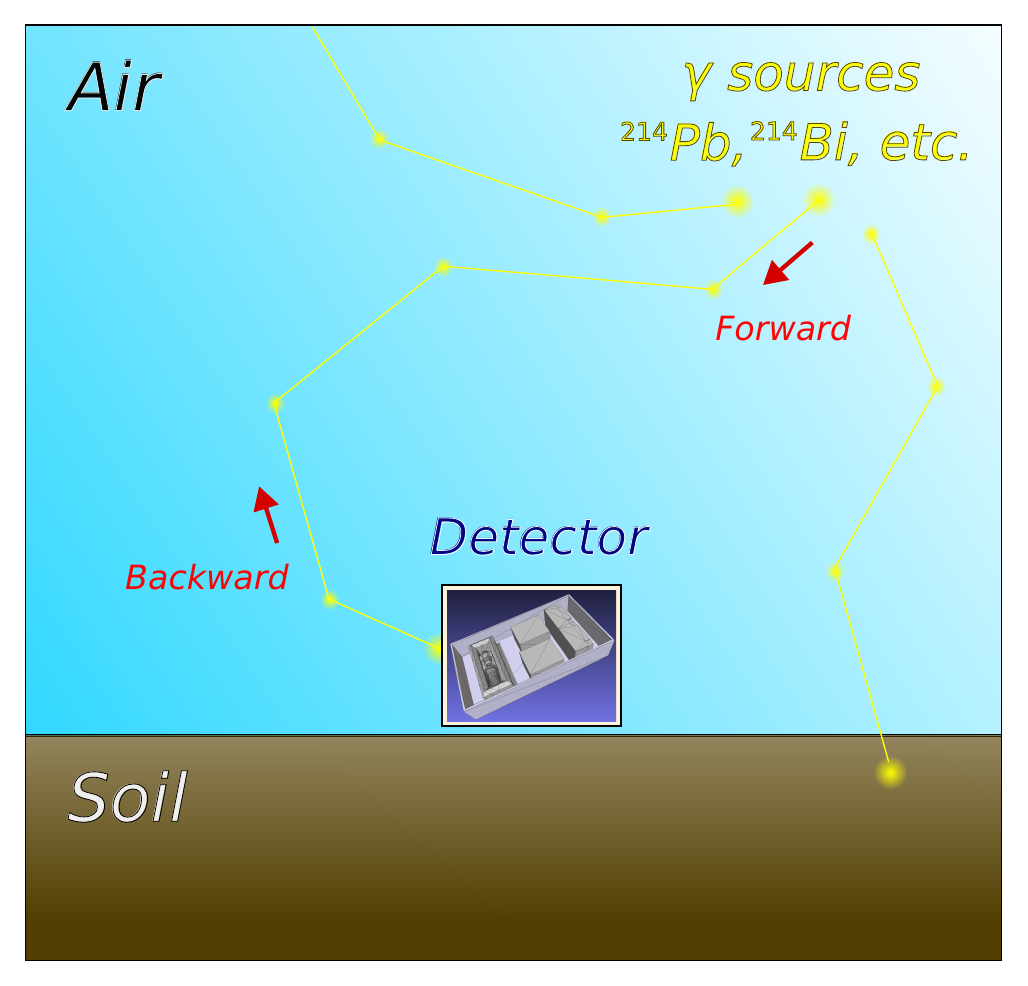}
    \caption{Schematic view of the Monte~Carlo transport of gamma photons
    emitted by radionuclides. Yellow lines represent Monte~Carlo trajectories,
    and yellow dots indicate collision vertices. Gamma photons follow
    rectilinear trajectories between two vertices.
    \label{fig:global-view}}
\end{figure}

During this exploratory work, we faced the problem of modelling the transport of
gamma photons from their emission regions to the spectrometer, which motivated
the present work. Properly modelling gamma-rays transport is crucial in order to
relate photon counts recorded in the spectrometer to concentrations of
radionuclides in the environment. But, before discussing the details of this
transport problem, let us first point out that we are only concerned with
low-energy gamma-rays emitted by radionuclides, i.e. with energies less than a
few MeV, as opposed e.g. to astrophysical gamma-rays which can reach energies of
the order of TeV (resulting in extensive air showers, which are not addressed
herein). Secondly, let us recall that, for most of the transport problem, gamma
photons can be considered as point-like particles originating from incoherent
sources and traveling in a straight line between two collisions with an atom of
the propagation medium, where its direction might change significantly (see e.g.
figure~\ref{fig:global-view}). Thirdly, note that a photon emitted at $1\,$MeV
undergoes $\mathcal{O}(10)$ collisions before being definitively absorbed (by
photoelectric effect).

The conventional approach to the present transport problem is to consider only
those photons that have not interacted between their points of emission and
detection, and that have been fully captured by the detector. This results in
selecting the peaks in the measured energy spectrum corresponding to the
emission lines of radioactive isotopes of interest. In this case, the transport
between the source and the detector can be considered as rectilinear. The photon
survival probability is calculated from the cross-section of all interaction
processes (absorption and scattering) that may occur along the way
(\citet{Minty1997}). This method, known as the point-kernel method, offers
simple calculations but at the price of sacrifying a large amount of data.
However, events statistics is often a limiting factor that one cannot afford to
sacrify.

In order to interpret full-spectrum data, i.e. including photo peaks and
background resulting from scattered photons, it is necessary to model in detail
the various interaction processes involved in the transport. For this purpose,
Monte~Carlo methods are commonly used due to the stochastic nature of collisions
between a gamma photon and an atom. Monte~Carlo methods represent the transport
problem as a discrete set of likely trajectories, known as Monte~Carlo
realisations. The classical approach is to generate Monte~Carlo realisations at
frequencies consistent with the physical reality. This approach is usually
called {\em analogue} Monte~Carlo. The Monte~Carlo method also allows for an
accurate modelling of the detector response, resulting in synthetic-like
measurements. Various codes, including
\Geant~\cite{Agostinelli2003,Allison2006,Allison2016} and \MCNP~\cite{MCNP2023},
have been employed for interpreting full-spectrum data (\citet{Bagatelas2010,
Androulakaki2016}).

Analogous Monte~Carlo methods are effective when the photon's source size is not
large as compared to the detector size (typically only ten centimetres). For
instance, this is the case in dense environments like water or soil, where the
range of gamma-rays is usually less than one metre. In contrast, the mean free
path between two collisions in air is large, approximately $100\,$m for a
$1\,$MeV photon. Therefore, a gamma-ray detector placed in the air is sensitive
to a large number of distant photons, but having a small probability of actually
reaching the spectrometer. Analogous Monte~Carlo methods require simulating all
these trajectories, which is particularly inefficient in the air, since most
simulated photons do not reach the detector.

Several solutions have been proposed to overcome this obstacle. For a
ground-atmosphere interface with an $(xOy)$ symmetry plane, and a gamma-ray
source that is also horizontally invariant, the photon flux depends only on
the altitude coordinate $z$. Therefore, it is possible to efficiently simulate
the gamma-ray flux at the observation altitude by using a fictitious detector
occupying an entire horizontal slice. Then, this flux can then be injected in a
realistic detector simulation (\citet{Baldoncini2018}). However, in practice,
assumptions of a flat geometry or a homogeneous source may be too strong
(see e.g. \citet{Satoh2014}), requiring alternative approaches.

Some authors have proposed mixed models that combine a Monte~Carlo simulation in
the vicinity of the detector with deterministic models of photon transport
beyond the detector (\citet{Smith2008, Shaver2009}). These deterministic models
are based on the resolution of a simplified radiative transfer equation.
%% However, it has been shown that these models have limitations in their
%% representation of Compton scattering.
%% XXX This paragraph might need some addition.

Another approach to address the Monte~Carlo inefficiency problem is to rely on
\acf{IS} methods. In particular, an efficient strategy is to reverse the
simulation of gamma-rays trajectories, starting from the detector until
incidentally reaching a source, as sketched in figure~\ref{fig:global-view}.
This is the approach adopted in this work. This method guarantees that all
simulated trajectories arrive at the detector, by construction. However, it does
not ensure that a backwards generated trajectory starts from a source of
gamma-rays. Nonetheless, when the source region is significantly larger than the
detector one, impressive improvements in simulation efficiency are obtained.

However, achieving accurate backward Monte~Carlo transport requires dedicated
algorithms and software developments that are technically more complex than
forward methods. Moreover, these functionalities are typically added to existing
forward Monte~Carlo software, not initially designed for running backwards,
which further complicates the task (see e.g.
\citet{Gabler2006,Pourrouquet2011,Robinson2022,Malins2017}, and
\citet{Desorgher2010} in the case of \Geant). Not all of these (modified)
software are publicly available. Furthermore, when available, comparisons to
forward results show significant discrepancies (see e.g.
\citet{Desorgher2010,Looper2018,Bongim2020}).

In addition, gamma-ray sources have a discrete emission spectrum, while existing
backwards going algorithms are typically designed to sample continuous spectra.
In the latter case, the stopping condition is determined solely by the position
of the particle, based on the geometry of the simulation, rather than its
energy. Thus, the final energy of a backwards generated trajectory is unlikely
to match any of the source emission lines. As a result, commonly available
backward Monte~Carlo implementations, such as the one used in \Geant, are not
suitable for our transport problem.

Recently, new techniques for backward transport have been developed
(\citet{Niess2018}), which efficiently account for the scattering of low-energy
muons in muography problems, with an accuracy comparable to forward methods (see
e.g. \citet{Niess2022}). These techniques are applied to gamma-ray transport,
which is discussed in section~\ref{sec:algorithms}. Additionally, we propose a
simple modification to the backwards going algorithm that enables sampling of
discrete emission lines. The modified algorithm has been implemented in a new
software library called \Goupil ({\bf G}amma transp{\bf O}rt {\bf U}tility,
a{\bf P}proximate but revers{\bf I}b{\bf L}e), which is dedicated to
transporting gamma-rays emitted by radionuclides. The presentation and
validation of this library are discussed in the rest of the paper, specifically
in sections~\ref{sec:implementation} and~\ref{sec:validation}.

\section{Transport and secondaries} \label{sec:particules-secondaires}

The algorithmic developments presented in this article were motivated by the
need to accelerate the atmospheric transport of gamma photons produced by
natural radioactivity in the air and soil. Radioactivity in the air comes mainly
from the short-lived daughters of $\ce{^{222}Rn}$, such as $\ce{^{214}Pb}$ and
$\ce{^{214}Bi}$. In the soil, it results from the decay chains of Uranium
($\ce{^{235}U}$, $\ce{^{238}U}$) and Thorium ($\ce{^{232}Th}$), as well as from
the decay of $\ce{^{40}K}$. These isotopes and their progenies are contained
in rocks. Therefore, a gamma photon detected at a particular observation point
is likely to originate from all directions in space. That is, the gamma photon
sources analysed in this study are diffuse in nature.

Furthermore, these sources exhibit a gamma emission spectrum that encompasses
various lines corresponding to the transition energies between the different
energy levels of the radionuclides in question. For all applications of the
methods presented, we use a simplified emission spectrum corresponding to the
primary emission lines accompanying the $\ce{^{214}Pb} \to \ce{^{214}Bi}$ and
$\ce{^{214}Bi} \to \ce{^{214}Po}$ decays (see table~\ref{tab:spectrum}). These
radionuclides are found both in the atmosphere and in the soil and cover an
energy range representing the gamma photons produced by natural and artificial
radioactivity. Nonetheless, the methods described below can also be applied to
other emission spectra.

\begin{table}[th]
    \caption{Main gamma emission lines in the $\ce{^{222}Rn}$ decay chain,
    according to~\cite{Lara}. Reported intensities are relative to the
    corresponding decay activity. Only emission lines with an intensity greater
    than $3\,\%$ are considered.
    \label{tab:spectrum}}
\center
    \begin{tabular}{lll}
\toprule
Decay & Energy (MeV) & Intensity ($\%$) \\
\midrule
$\ce{^{214}Pb} \to \ce{^{214}Bi}$ & $0.242$ & $   7.3$ \\
& $0.295$ & $  18.4$ \\
& $0.352$ & $  35.6$ \\
\midrule
$\ce{^{214}Bi} \to \ce{^{214}Po}$ & $0.609$ & $  45.5$ \\
& $0.768$ & $   4.9$ \\
& $0.934$ & $   3.1$ \\
& $1.120$ & $  14.9$ \\
& $1.238$ & $   5.8$ \\
& $1.378$ & $   4.0$ \\
& $1.764$ & $  15.3$ \\
& $2.204$ & $   4.9$ \\
\bottomrule
\end{tabular}

\end{table}

\begin{figure}[th]
    \center
    \includegraphics[width=\textwidth]{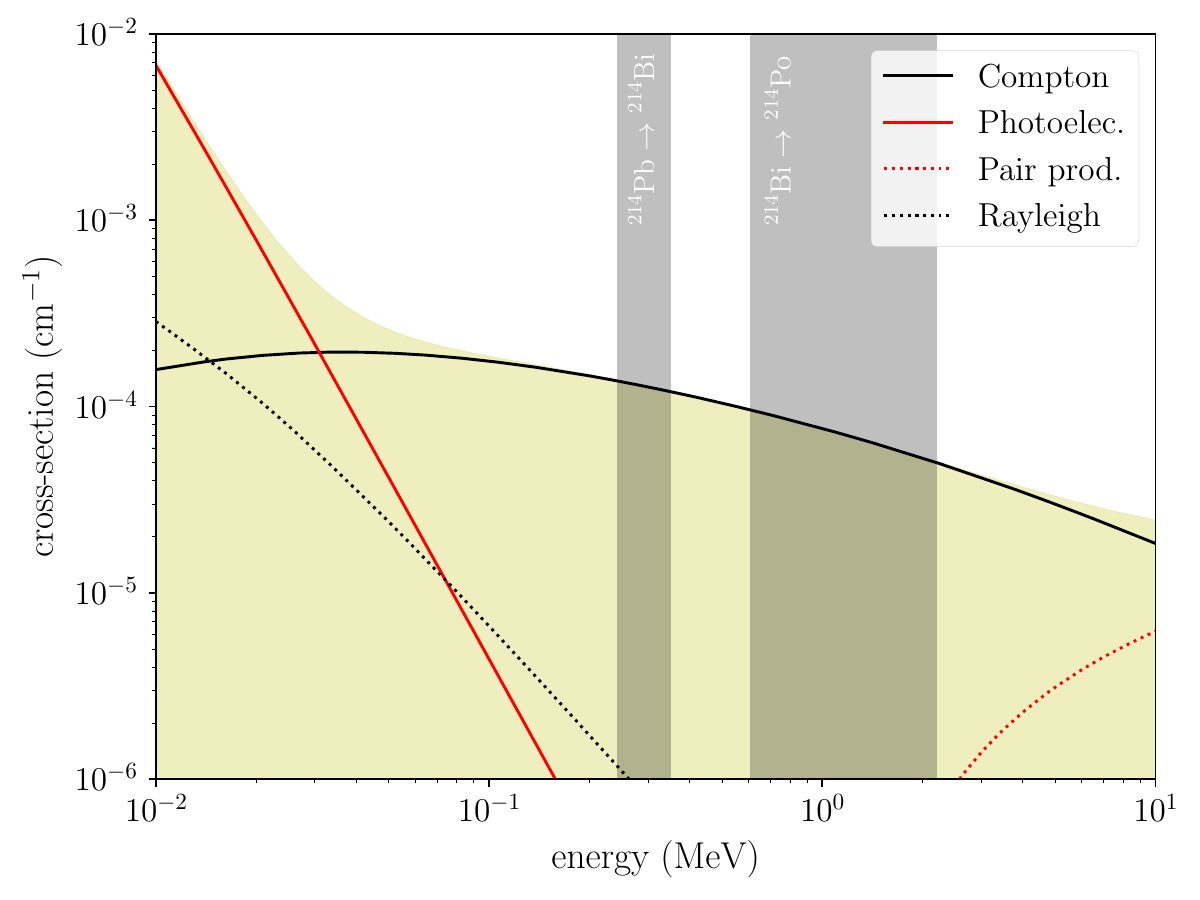}
    \caption{Macroscopic cross-sections for photon interactions in dry air at 1
    bar ($\rho = 1.205\,$mg$/$cm$^3$), as per the EPDL (\citet{Cullen1997}). The
    total cross-section is represented by the yellow shaded area, which is
    obtained by summing the four processes shown in the figure. The grey
    vertical bands indicate the energy ranges of the gamma emissions associated
    with the $\ce{^{214}Pb}\to\ce{^{214}Bi}$ and $\ce{^{214}Bi}\to\ce{^{214}Po}$
    decays, according to table~\ref{tab:spectrum}.
    \label{fig:cross-section}}
\end{figure}

Figure~\ref{fig:cross-section} displays the cross-sections of the various
interaction mechanisms between a gamma photon and air molecules. Gamma photons
lose their energy primarily through Compton collisions, which are dominant at
emission energies. Eventually, photons are absorbed by the photoelectric effect
when their energy decreases to less than a hundred keV. These processes release
a secondary electron that can also propagate. In addition, for energies greater
than $2m_e \simeq 1\,$MeV (where $m_e$ is the mass of the electron), photons can
be converted into an $e^+e^-$ pair within the electromagnetic field of medium's
atoms (pair production). Moreover, gamma-rays may be emitted by secondary
electrons and positrons through braking radiation (bremsstrahlung) or
electron-positron annihilation. To determine the total flux of particles that
ultimately reach the detector, one should in principle consider the entire chain
of interactions, including primary and secondary particles.

To quantify the secondary flux, we conducted a synthetic experiment using \Geant
(version 10.7.4, \texttt{G4EmStandardPhysics}). We used an isotropic and
homogeneous propagation medium filled with dry air of density $1.205\,$mg/cm$^3$,
which corresponds to temperature conditions of $20\,^\circ$C under $1\,$bar of
atmospheric pressure. A point source of gamma-rays is placed in this medium,
which is isotropic in direction and has an emission spectrum consistent with the
progeny of $\ce{^{222}Rn}$, according to table~\ref{tab:spectrum}. The
experiment is to count the outgoing particles through a fictitious sphere of
radius $r$ centred on the source.

\begin{figure}[th]
    \center
    \includegraphics[width=\textwidth]{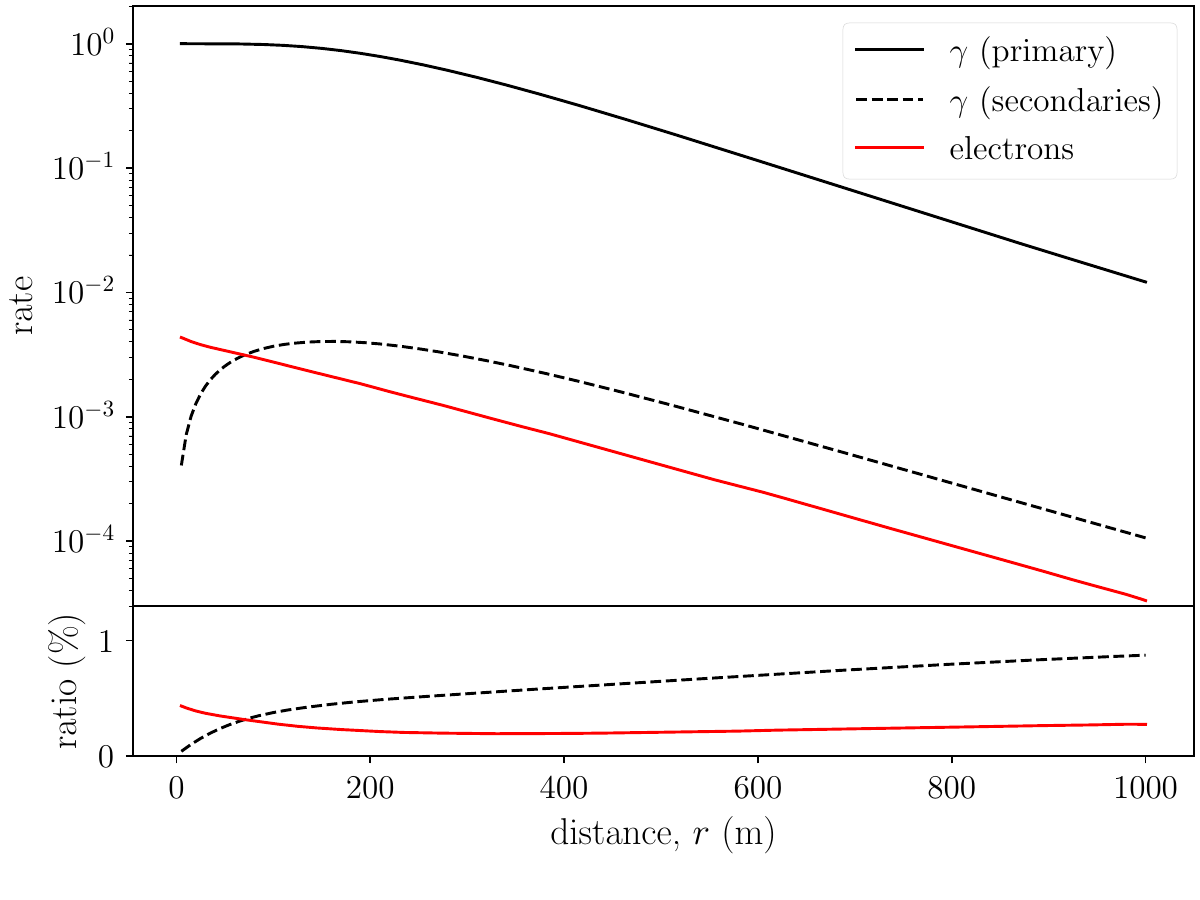}
    \caption{Normalised rate of outgoing particles at a distance $r$ from the
    source for the synthetic experiment described in the text (\Geant 10.7.4,
    \texttt{G4EmStandardPhysics}).  The lower frame displays the ratio to
    primary photons as a percentage. The positron flux is negligible compared to
    that of the electrons and secondary photons.
    \label{fig:flux-experiment}}
\end{figure}

Figure~\ref{fig:flux-experiment} shows the results obtained at different
distances, $r$, from the source. The outgoing flux is composed of less than
$1\,\%$ secondary particles. This is due to the fact that electrons are less
penetrating than photons, and that the probability of emitting a secondary gamma
photon is low, at these energies. Therefore, secondary particles have a minimal
contribution to the transport. However, secondary particles account for $96\,\%$
of the CPU computing time.

Therefore, we will only transport primary photons in order to improve
computational efficiency (which results in a slight underestimation of gamma-ray
fluxes, up to one percent). In addition, this approximation
significantly simplifies the backward transport, discussed hereafter.
Nonetheless, it is important to remember that secondary particles are crucial in
simulating detector response, as they are responsible for depositing energy in
sensitive materials. Neglecting them is only possible at a
sufficient distance from the detector's sensitive zones. What constitutes a
sufficient distance depends on the application and the desired accuracy. The
particular case of~a~scintillation detector is discussed in the validation
section (\autoref{sec:test2}).

\section{Transport algorithms \label{sec:algorithms}}

The details of the Monte~Carlo transport of gamma-rays are discussed below. For
the sake of clarity, we will first consider a simple setup that captures the
essence of the problem. We will then present a forward transport algorithm and
its inversion. Finally, we will discuss the generalisation to more realistic
setups.

\subsection{Problem formulation}

A set of gamma-ray sources is assumed to be distributed within a closed
propagation medium bounded by an external contour $\mathcal{E}$. Inside this
medium, a second closed volume containing the detector is isolated. The contour
of this second volume is called the collection surface, $\mathcal{C}$. In
addition, it is assumed that there are no sources below $\mathcal{C}$. The
Monte~Carlo transport amounts to generating a set of trajectories from the
sources to a point in $\mathcal{C}$, without intersecting $\mathcal{E}$, and
according to the laws of physics. The set of these trajectories is denoted $T$.
It is further required that $T$-trajectories intersect $\mathcal{C}$ exactly
once, at the end. This condition will become useful when the problem is
generalised in section~\ref{sec:generalisation}. For the time being, note that
it amounts to stopping the simulation as soon as a trajectory intersects with
$\mathcal{C}$.

The trajectory of a Monte~Carlo photon consists in a sequence of $n$ collision
vertices connected by straight-line segments, as illustrated in
figure~\ref{fig:schema-trajectoire} below. In addition, the trajectory includes
an incoming segment from the source to the first collision and an outgoing
segment from the last collision to the collection surface. The positions of the
collision vertices are denoted as $\vb{r}_i$ with $i \in [1, n]$, and by
extension $\vb{r}_0 = \vb{r}_I$ ($\vb{r}_{n+1} = \vb{r}_F$) represents the
starting (arrival) point of the trajectory. The photon momentum at the exit of
$\vb{r}_i$ is denoted $\vb{k}_i = \nu_i \vb{u}_i$, where $\nu_i$ is the photon
energy and $\vb{u}_i$ a unit vector representing its momentum direction. Note
that the momentum remains constant along a segment, e.g. between two collisions.
Thus, a Monte~Carlo trajectory is represented by an ordered set of states, $S_i
= \{\nu_i, \vb{r}_i, \vb{u}_i\}$.

\begin{figure}[th]
    \center
    \includegraphics[width=\textwidth]{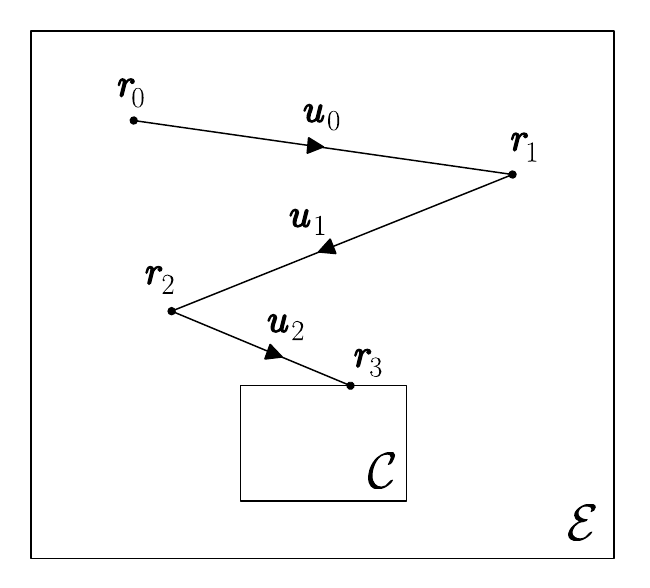}
    \caption{Example of a 2D Monte~Carlo trajectory that was produced at point
    $\vb{r}_0$. The trajectory reaches the collection surface $\mathcal{C}$ at
    point $\vb{r}_3$ after two collisions, at $\vb{r}_1$ and $\vb{r}_2$. The
    collection surface ($\mathcal{C}$) and the external surface ($\mathcal{E}$)
    are represented by rectangles.
    \label{fig:schema-trajectoire}
}
\end{figure}

It is assumed that the propagation medium has a spatially uniform composition,
but a potentially variable density $\rho$, e.g. representing the Earth's
atmosphere. Therefore, the atomic cross-section $\sigma$, for photons
interactions with atoms of the propagation medium, depends solely on $\nu$, not
on $\vb{r}$, nor on $\vb{u}$. However, the mean free path of photons depends on
the density $\rho$ of the medium, and therefore on $\vb{r}$, as
\begin{equation}
    \frac{1}{\lambda(\nu, \vb{r})} = \rho(\vb{r}) \frac{\mathcal{N}_A}{M}
        \sigma(\nu),
\end{equation}
where $M$ represents the molar mass of the propagation medium, which is constant
based on our assumptions. Thus, in some circumstances it is relevant to instead
consider the mean free path in units of grammage, defined as
\begin{equation}
    \Lambda(\nu) = \rho \lambda,
\end{equation}
which only depends on $\nu$.

To account for the amount of target material between two vertices in a medium of
variable density, let us define the grammage function $X(d; \vb{r}, \vb{u})$, as
\begin{equation}
    X(d; \vb{r}, \vb{u}) = \int_0^{d}{\rho(\vb{r} + s \vb{u}) ds} ,
\end{equation}
where $d$ represents the distance travelled from $\vb{r}$ along the line of
sight $\vb{u}$. We denote $d_i = \|\vb{r}_i - \vb{r}_{i-1}\|$ the distance
between vertices $i - 1$ and $i$, and $X_i = X(d_i; \vb{r}_{i-1}, \vb{u}_{i-1})$
the column depth (grammage) between these same vertices. Note that for $\rho >
0$, the grammage function is always monotonically increasing according to $d$,
making it invertible given $\vb{r}$ and $\vb{u}$.

The survival probability of a photon between vertices $i$ and $i+1$ depends on
$X_i$ as
\begin{equation} \label{eq:proba-de-survie}
    P_{s, i} = e^{-\ell_i}, \quad
    \ell_i   = \frac{X_i}{\Lambda(\nu_{i-1})} .
\end{equation}
The corresponding \ac{PDF} of interacting at the distance $d_i$ from
$\vb{r}_{i-1}$ is
\begin{equation} \label{eq:proba-interaction}
    p_{s, i} = \frac{P_{s,i}}{\lambda(\nu_{i-1}, \vb{r}_i)} .
\end{equation}

At a collision vertex, the energy of the photon changes discontinuously from
$\nu_{i-1}$ to $\nu_i$, and its direction changes from $\vb{u}_{i-1}$ to
$\vb{u}_i$. To account for the different dynamics of photon interaction
processes, let us write the corresponding collision \ac{PDF} as a mixture
density, as
\begin{equation}
    \label{eq:proba-de-collision}
\begin{aligned}
    p_c\left(\nu_i, \vb{u}_i; \nu_{i-1}, \vb{u}_{i-1}\right) &={} \sum_j {
        q_j\left(\nu_i; \nu_{i-1}\right)
        p_{c,j}\left(\nu_i, \vb{u}_i; \nu_{i-1}, \vb{u}_{i-1}\right)
    }, \\
    q_j\left(\nu_i; \nu_{i-1}\right) &={}
        \frac{\sigma_j(\nu_{i-1})}{\sigma(\nu_{i-1})}, \\
    p_{c,j}\left(\nu_i, \vb{u}_i; \nu_{i-1}, \vb{u}_{i-1}\right) &={}
        \frac{1}{\sigma_j(\nu_{i-1})}
        \frac{\partial^2\sigma_j}{\partial\nu_i \partial\Omega_i}, \\
\end{aligned}
\end{equation}
where  $\Omega_i$ represents the solid angle at which the photon emerges during
a collision. The index $j$ runs over interaction processes, such that $\sigma =
\sum\sigma_j$. The details of these processes, such as their cross-sections, are
not relevant at this point, and will be discussed in section \ref{sec:physique}.

A Monte~Carlo trajectory is characterised by a sequence of states, $\vb{S}$,
resulting from a succession of statistically independent random events, as
follows: first, interacting at the point $\vb{r}_i$ (event of density $p_{s,i}$,
\autoref{eq:proba-interaction}); then, resolving this interaction (event of
density $p_c$, \autoref{eq:proba-de-collision}); and so on. The \ac{PDF}
$p(\vb{S})$ of a trajectory corresponding to this succession of events is
important for the rest of the discussion. This \ac{PDF} can be expressed from
the previous definitions. Note that the last step of the trajectory,
intercepting the collection surface, differs from the others. In this case, it
is sufficient to consider the survival probability $P_{s,n+1}$ of the photon
between vertices $n$ and $n+1$ (\autoref{eq:proba-de-survie}). After a few
simplifications, $p(\vb{S})$ writes
\begin{equation} \label{eq:proba-trajectoire}
    p(\vb{S}) =
        P_s(\vb{S}) c(\vb{S}), \quad
        c(\vb{S}) = \prod_{i=1}^{n}{
            \frac{p_{c,\vb{j}(i)}\left(
                \nu_i, \vb{u}_i; \nu_{i-1}, \vb{u}_{i-1}\right)}{
                \lambda_{\vb{j}(i)}(\vb{r_i, \nu_{i-1}})}
        },
\end{equation}
where $\vb{j}$ is an index vector indicating the interaction processes for the
$n$ collision vertices. The prefactor $P_s$ corresponds to the survival
probability of the photon over the entire trajectory, i.e. between the emission
point and the collection point. This survival probability writes
\begin{equation} \label{eq:proba-survie-totale}
    P_s = \exp\left(-\sum_{i=1}^{n+1}{\ell_i}\right) .
\end{equation}

Finally, when discussing the following transport algorithms, it is useful to
distinguish between three types of interaction processes: absorption ($\nu_{i+1}
= 0$), elastic ($\nu_{i+1} = \nu_i$) and inelastic ($\nu_i > \nu_{i+1} > 0$).
The corresponding sets of $j$ process indices are denoted as
$\textproc{Absorption}$, $\textproc{Elastic}$, and $\textproc{Inelastic}$.

\subsection{Forward algorithm \label{sec:direct-algorithm}}

With the previous notations and definitions, we can now discuss the forward
algorithm for generating Monte~Carlo trajectories, detailed below
(\autoref{al:generation-directe}). This is a classical algorithm (see e.g.
\citet{Berger1963}). But, we find useful to recall it explicitly for the sake of
understanding of the reverse method.

Some steps in algorithm~\ref{al:generation-directe} are represented by external
functions that are not described herein, for brevity. For instance, the function
\textproc{Open01} (\autoref{line:open01}) consumes a random stream $\mathcal{R}$
and returns a real number that is uniformly distributed over the open interval
$(0, 1)$. By using the open interval instead of $[0, 1]$, we escape some
numerical issues. Let us also point out that the function
\textproc{CrossSections} (\autoref{line:cross-sections-1},
\autoref{line:cross-sections-2}) returns a vector of values that correspond to
the cross-sections of the various interaction processes.

The functions \textproc{DistanceToVertex} (\autoref{line:distance-to-vertex})
and \textproc{DistanceToInterface} (\autoref{line:distance-to-interface})
account for the geometry of the propagation medium. The first function inverts
the column depth (grammage, $X$) along the line of sight, $\vb{u}$, to determine
the geometric distance $d$ (noted $d_V$ in
algorithm~\ref{al:generation-directe}). In the case of a medium with uniform
density, this inversion is straightforward, as $d = X / \rho$. However, in
general, this step might be complex, although always solvable for $\rho > 0$, as
previously noted. The second function, \textproc{DistanceToInterface}, returns
the distance to the first intersected interface along a line of sight. This
function may be provided by an external ray-tracing algorithm. For the simple
setup considered herein, the intersected interface is either the collection
surface $\mathcal{C}$ or the external border $\mathcal{E}$ encapsulating the
source medium. In the absence of a boundary, function
\textproc{DistanceToInterface} would return $+\infty$, for example.

The two other functions appearing in algorithm~\ref{al:generation-directe}
simulate collisions using the random stream $\mathcal{R}$. Function
\textproc{SelectProcess} (\autoref{line:select-process}) is rather simple as it
selects one of the interaction processes with probability $q_j = \sigma_j /
\sigma$. However, an efficient implementation of function
\textproc{RandomiseCollision} (\autoref{line:randomise-collision-1},
\autoref{line:randomise-collision-2}) is more complex, which will be discussed
in section \ref{sec:implementation}.

\begin{algorithm}[h]
    \caption{
Forward Monte~Carlo transport.\label{al:generation-directe}
}
    \begin{algorithmic}[1]
        \Require $\nu_I > 0$, $\|\vb{u}_I\| = 1$,
        $\mathcal{R}$ (a random stream).
    \State ($\nu, \vb{r}, \vb{u}) \gets (\nu_I, \vb{r}_I, \vb{u}_I)$
        \Comment{Initialisation to source parameters.}
    \State $\vb*{\sigma} \gets \Call{CrossSections}{\nu}$
        \label{line:cross-sections-1}
    \State $\sigma \gets \Call{Sum}{\vb*{\sigma}}$
    \Loop
        \State $\Lambda \gets \frac{M}{\mathcal{N}_A \sigma}$
        \State $\zeta \gets \Call{Open01}{\mathcal{R}}$
            \label{line:open01}
        \State $X \gets -\Lambda \ln \zeta$
        \State $d_V \gets \Call{DistanceToVertex}{\vb{r}, \vb{u}, X}$
            \label{line:distance-to-vertex}
        \State $d_I \gets \Call{DistanceToInterface}{\vb{r}, \vb{u}}$
            \label{line:distance-to-interface}
        \If{$d_I < d_V$} \label{line:arret-1}
            \Comment{Stop on an interface.}
            \State $\vb{r} \gets \vb{r} + d_I \vb{u}$
            \State \Return{($\nu, \vb{r}, \vb{u})$}
                \label{line:return-1}
        \Else
            \State $\vb{r} \gets \vb{r} + d_V \vb{u}$
        \EndIf
        \State $j \gets \Call{SelectProcess}{\vb*{\sigma}, \sigma,
            \mathcal{R}}$ \label{line:select-process}
        \If{$j \in \textproc{Absorption}}$ \label{line:arret-2}
            \Comment{Stop by absorption.}
            \State \Return{($0, \vb{r}, \vb{u})$}
        \ElsIf{$j \in \textproc{Elastic}}$
            \State $(\cdot, \vb{u}) \gets \Call{RandomiseCollision}{
                j, \nu, \vb{u}, \mathcal{R}}$ \label{line:randomise-collision-1}
        \Else \label{line:after-else}
            \State $(\nu, \vb{u}) \gets \Call{RandomiseCollision}{
                j, \nu, \vb{u}, \mathcal{R}}$ \label{line:randomise-collision-2}
            \State $\vb*{\sigma} \gets \Call{CrossSections}{\nu}$
                \label{line:cross-sections-2}
            \State $\sigma \gets \Call{Sum}{\vb*{\sigma}}$
        \EndIf
    \EndLoop
\end{algorithmic}
\end{algorithm}

Let us also emphasize some technical properties of
algorithm~\ref{al:generation-directe}. First, this algorithm generates two
trajectory topologies, depending on the stop condition (\autoref{line:arret-1}
or \autoref{line:arret-2}). The first topology, which stops on an interface,
contains the $T$-trajectories of interest. Note that in this topology, we also
find trajectories leaving the propagation medium through $\mathcal{E}$. The
second topology contains trajectories that are absorbed before ever reaching an
interface.

Secondly, algorithm~\ref{al:generation-directe} is analogue, i.e. trajectories
are generated according to $p(\vb{S})$ (\autoref{eq:proba-trajectoire}).
Therefore, in the following we equate {\em forward} and {\em analogue}. Note
however that, generally speaking, forward Monte~Carlo procedures might rely on
Importance Sampling methods, contrary to algorithm~\ref{al:generation-directe}.

\subsection{Backward algorithm}

As backward Monte~Carlo methods are uncommon, we shall first discuss some of
their specificities before proceeding with the inversion of the forward
procedure (\autoref{al:generation-directe}).

\subsubsection{Backward Monte~Carlo methods}

A Monte~Carlo transport procedure defines a functional relationship between the
departure state $S_I$ and the arrival state $S_F$ of a trajectory. This
relationship can be expressed as
\begin{equation}
    S_F = \textproc{F}(S_I, \mathcal{R}),
\end{equation}
where $\mathcal{R}$ represents a random stream that is consumed by
\textproc{F}.
Let us further assume that the functional inverse according to $S$ of
\textproc{F} exists, such that
\begin{equation}
    S_I = \textproc{B}(S_F, \mathcal{R}) .
\end{equation}
Then, it is important to note that the composition
\begin{equation}
    \textproc{B}(\cdot, \mathcal{R}') \circ
        \textproc{F}(\cdot, \mathcal{R}) (S_I) \neq S_I ,
\end{equation}
does not restore the initial state $S_I$. This is because the outcomes of the
\textproc{B} and \textproc{F} transport procedures depend on statistically
independent random streams, $\mathcal{R}'$ and $\mathcal{R}$. That is, backward
Monte~Carlo transport does not reverse time in the same way as rewinding a film.
Instead, the forward and backward procedures both reflect the collision
physics, which is inherently stochastic.

It is also interesting to observe that if the forward procedure is analogue,
usually the inverse one is not. Indeed, let us denote $p_F$ ($p_B$) the \ac{PDF}
of producing a trajectory with procedure \textproc{F} (\textproc{B}). Then, for
a trajectory $\vb{S}$, we have~\cite{Niess2018}
\begin{equation}
    p_B(\vb{S}) dS_I = p_F(\vb{S}) dS_F .
\end{equation}
Thus, the two procedures cannot be simultaneously analogue, except for unitary
processes, e.g. that conserve the particle's energy.

The functional inverse $\textproc{B}$ is a particular case of a backward
procedure, which is not necessarily optimal and its existence is not guaranteed
either. In practice, one frequently relies on alternative methods, such as an
adjoint formulation or the use of approximate processes (see e.g. corollary 1 of
\citet{Niess2018}). In any case, the physical correctness is restored by
weighting backwards simulated trajectories by a factor $\omega$, such that
\begin{equation} \label{eq:ponderation-trajectoire}
    \omega(\vb{S}) p^*(\vb{S}) = p(\vb{S}),
\end{equation}
where $p^*$ denotes the \ac{PDF} of the backward generation procedure, whether
the inversion is functional or not.

Consequently, backward algorithms belong to the family of \acf{IS} methods. A
valid backward algorithm produces any trajectory of interest generated by the
forward algorithm (i.e. $p^*(\vb{S}) > 0$ for all $\vb{S} \in T$), and the
corresponding $\omega$ weights satisfy to
equation~\ref{eq:ponderation-trajectoire}. The efficiency of the backward
method, i.e. Monte~Carlo convergence, depends on the distribution of
weights~$\omega$. It is advisable to achieve a relatively uniform distribution
of weights, as a finite number of Monte Carlo samples are used in practice. With
that respect, let us point out that in the case of functional inversion, the
weights are given by the Jacobian factor $\omega = \left|d\nu_I /
d\nu_F\right|$, which tends to be $\mathcal{O}(1)$ for non-chaotic systems.

\subsubsection{Constrained backward collisions}

After these initial comments, let us now consider the inversion of
algorithm~\ref{al:generation-directe} itself. Essentially, this algorithm can be
inverted by applying corollary~3 of \citet{Niess2018}. The main difficulty lies
in the backward simulation of collisions. With usual methods, the photon's
initial energy ($\nu_{i-1}$) is not restricted by the source energy ($\nu_I$),
but only by the kinematics of the physical process. To account for this external
constraint, a simple yet efficient solution is to override the energy
$\nu_{i-1}$ of the incoming photon to $\nu_I$ for any collision where $\nu_{i-1}
\geq \nu_I$. In addition, a corrective weight must be applied whenever
$\nu_{i-1}$ is overridden, as
\begin{equation}
    \omega(\nu_i, \nu_{i-1}) = \frac{p_c(\nu_i; \nu_{i-1})}
        {1 - P^*_c(\nu_{i-1}; \nu_i)} ,
\end{equation}
where $p_c$ is the \ac{PDF} for a forward collision, and $P^*_c$ the \ac{CDF}
for an unconstrained backward collision. This procedure is essentially similar
to stopping a particle as soon as it crosses a boundary surface, but operating
on the particle's energy instead of its position. The corresponding
algorithm~\ref{al:contrainte-ajustée} is detailed below, and a proof is provided
in \ref{sec:proof-constrained}.

\begin{algorithm}[h]
    \caption{Constrained Backward Collision.
        \label{al:contrainte-ajustée}}
    \begin{algorithmic}[1]
    \Require $\nu_i \in (0, \nu_I)$, $\mathcal{R}$ (a random stream).

        \State $(\nu_{i-1}, \omega_c) = \Call{RandomiseBackward}{
            \nu_i, \mathcal{R}}$
        \If{$\nu_{i-1} < \nu_I$}
            \State \Return $(\nu_{i-1}, \omega_c)$
        \Else
            \State $p_c \gets \Call{PdfForward}{\nu_i, \nu_I}$
                \label{line:densité-directe}
            \State $P^*_c \gets \Call{CdfBackward}{\nu_I, \nu_i}$
                \label{line:densité-cumulée-inverse}
            \State $\omega_c \gets \frac{p_c}{1 - P^*_c}$
            \State \Return $(\nu_I, \omega_c)$
        \EndIf
    \end{algorithmic}
\end{algorithm}

\subsubsection{Backward absorption} \label{sec:absorption-inverse}

Another point requiring attention is the backward treatment of absorption
processes. Let us recall that, as discussed in
section~\ref{sec:particules-secondaires}, we only consider primary photons. This
means that any collision process that does not restore the incident photon is
considered as absorbing and terminates the Monte~Carlo trajectory. But, in the
case of a backward collision, the absorbing processes are undetermined as we do
not specify any of the collision's product. Despite this, we cannot disregard
them as they alter the survival probability $P_s$ of $T$-trajectories, which
appears in equation~\ref{eq:proba-trajectoire} and gives $p$.

One possible solution is to treat absorption as a continuous process, as
demonstrated in \ref{sec:continuous-absorption}. Let $\Lambda_a$ be the mean
free path of the absorption processes only. During trajectory randomisation,
absorption processes are disregarded, but they are taken into account with an
additional weight of $\omega_a = P_s[\Lambda_a]$, which corresponds to the
trajectory survival probability according to the absorption processes alone
(i.e. \autoref{eq:proba-de-survie}, but with $\Lambda_a$ substituted for
$\Lambda$).

A more subtle alternative is to modify the \ac{PDF} of backwards generated
trajectories by a factor of $P_s[\Lambda_a] \in [0,1]$, instead of modifying
their weights. This can be achieved by randomly eliminating backward
trajectories with a probability of $r = 1 - P_s[\Lambda_a]$ using a Russian
Roulette method. Moreover, doing this step-by-step, between two successive
vertices, avoids unnecessary resource wastage in generating full trajectories
that would eventually be eliminated. Section~\ref{sec:algorithme-complet}
hereafter presents a complete backward algorithm using the latter method.

\subsubsection{Backward photo-peaks} \label{sec:backward-photopeaks}

Before discussing the complete backward transport, we shall consider the
particular case of photo-peak trajectories. These events are elastic, meaning
that the photon energy, $\nu = \nu_I$, is conserved over the entire trajectory.
However, the photon may still undergo elastic collisions, such as Rayleigh
scattering. As elastic collisions are self-inverse processes (see e.g.
Appendix~D of \citet{Niess2018}), the forward and backward transport of
photo-peak events can be made almost identical. The only difference lies in the
stopping condition. In the forward case, photo-peak trajectories stop when they
intersect the collection surface $\mathcal{C}$. In the backward case, the
forward algorithm~\ref{al:generation-directe} can be used (starting from $S_F$),
but the trajectory should be stopped as soon as an inelastic collision would
occur (i.e. after \autoref{line:after-else}, before simulating the inelastic
collision). The corresponding stop point, $\vb{r}_I$, is selected as a potential
source location for the backward trajectory. Furthermore, the backward
trajectory should be weighted by a factor
\begin{equation}
    \omega = \lambda_\text{in}(\vb{r}_I, \nu_I),
\end{equation}
where $\lambda_\text{in}$ represents the mean free path for inelastic processes
only, excluding absorbing and elastic processes. A proof of this method is
provided in \ref{sec:photopeaks-proof}. Note that this approach considers
absorption processes as self-inverses, which is consistent with the previous
discussion (\autoref{sec:absorption-inverse}).

\subsubsection{Complete backward algorithm} \label{sec:algorithme-complet}

Using the constrained backward collision procedure,
\textproc{RandomiseConstrained}, as previously stipulated by
algorithm~\ref{al:contrainte-ajustée}, we can invert
algorithm~\ref{al:generation-directe}.  Applying corollary~3 of
\citet{Niess2018}, in the absence of any continuous process, leads to the
algorithm~\ref{al:generation-inverse} below, where the weight $\omega_c$
(\autoref{line:randomise-constrained}, \autoref{line:poids-collision})
corresponds to the term $\left|\partial\epsilon_{j+1}/\partial\epsilon_j\right|
\omega_b$ of \cite[eq.~15]{Niess2018} (using the original notations). However,
note that $T$-trajectories are terminated by the condition of
line~\autoref{line:arret-source} instead of an interface crossing. This
corresponds to the specific case of photo-peaks transport, which was discussed
previously in section~\ref{sec:backward-photopeaks}. The stopping condition is
reached either by setting $\nu_F = \nu_I$ from the beginning or when an
inelastic collision triggers the energy constraint $\nu_I$.

Let us point out that the sampling of the interaction process
(\autoref{line:select-process-inverse}) uses the same physical cross-sections as
in the forward case, due to Lemma~5 of \cite{Niess2018}.  This enables the use
of the same backward absorption process as in the forward case. However, Lemma~5
requires $\sigma_j > 0$ for all $j$, which is not verified in this case for the
production of an $e^+e^-$ pair. This is kinematically forbidden for $\nu < 2
m_e$. One possible solution is to group this process with the photoelectric
interaction. Both processes are of absorption type and therefore identical from
the point of view of the transport algorithm. This approach ensures that the
total cross-section of the absorption processes verifies $\sigma_a > 0$.

\begin{algorithm}[h]
    \caption{
Backward Monte~Carlo transport.\label{al:generation-inverse}
}
    \begin{algorithmic}[1]
        \Require $\nu_I \geq \nu_{n+1} > 0$, $\|\vb{u}\| = 1$,
            $\sigma_j > 0\,\forall j$,
            $\mathcal{R}$ (a random stream).
    \State $(\nu, \vb{r}, \vb{u}) \gets
        (\nu_{n+1}, \vb{r}_{n+1}, \vb{u}_{n+1})$
        \Comment{Initialisation to arrival parameters.}
    \State $\omega \gets 1$
    \State $\vb*{\sigma} \gets \Call{CrossSections}{\nu}$
    \State $\sigma \gets \Call{Sum}{\vb*{\sigma}}$
    \Loop
        \State $\Lambda \gets \frac{M}{\mathcal{N}_A \sigma}$
        \State $\zeta \gets \Call{Open01}{\mathcal{R}}$
        \State $X \gets -\Lambda \ln \zeta$
        \State $d_V \gets \Call{DistanceToVertex}{\vb{r}, -\vb{u}, X}$
        \State $d_I \gets \Call{DistanceToInterface}{\vb{r}, -\vb{u}}$
        \If{$d_I < d_V$} \label{line:arret-interface-inverse}
            \Comment{Stop on an interface.}
            \State $\vb{r} \gets \vb{r} - d_I \vb{u}$
            \State \Return{($\nu, \vb{r}, \vb{u}, \omega)$}
        \Else
            \State $\vb{r} \gets \vb{r} - d_V \vb{u}$
        \EndIf
        \State $j \gets \Call{SelectProcess}{\vb*{\sigma}, \sigma,
            \mathcal{R}}$ \label{line:select-process-inverse}
        \If{$j \in \textproc{Absorption}}$ \label{line:arret-absorption-inverse}
            \Comment{Stop by absorption.}
            \State \Return{($0, \vb{r}, \vb{u}, 0)$}
        \ElsIf{$j \in \textproc{Elastic}$}
            \State $(\cdot, \vb{u}) \gets \Call{RandomiseCollision}{
                j, \nu, \vb{u}, \mathcal{R}}$
        \ElsIf{$\nu = \nu_I$} \label{line:arret-source}
            \Comment{Stop on a source.}
            \State $\rho \gets \Call{Density}{\vb{r}}$
            \State $\sigma_\text{in} \gets
                \Call{Sum}{\vb*{\sigma}(\textproc{Inelastic})}$
            \State $\lambda_\text{in} \gets \frac{M}{\rho\,\mathcal{N}_A\sigma_\text{in}}$
            \State $\omega \gets \lambda_\text{in}\, \omega$ \label{line:correction-vertex}
            \State \Return{($\nu, \vb{r}, \vb{u}, \omega)$}
        \Else
            \State $(\nu, \vb{u}, \omega_c) \gets \Call{RandomiseConstrained}{
                j, \nu, \vb{u}, \nu_I, \mathcal{R}}$
                \label{line:randomise-constrained}
            \State $\sigma_j \gets \vb*{\sigma}(j)$
            \State $\vb*{\sigma} \gets \Call{CrossSections}{\nu}$
            \State $\sigma \gets \Call{Sum}{\vb*{\sigma}}$
            \State $\omega \gets \omega_c \frac{\vb*{\sigma}(j)}{\sigma_j}
                \omega$ \label{line:poids-collision}
        \EndIf
    \EndLoop
\end{algorithmic}
\end{algorithm}

As a cross-check, it is useful to determine the \ac{PDF} of the $T$-trajectories
generated by algorithm~\ref{al:generation-inverse}.  After simplification, we
find
\begin{equation}\label{eq:proba-trajectoire-inverse}
\begin{aligned}
    p^*(\vb{S}) &={}
        P_s(\vb{S}) c^*(\vb{S}), \\
    c^*(\vb{S}) &={} \frac{1}{\lambda_\text{in}(\vb{r}_I, \nu_I)}
        \prod_{i=1}^{n}{
            \frac{p_{c,\vb{j}(i)}^*
                \left(\nu_{i-1}, \vb{u}_{i-1}; \nu_i, \vb{u}_i\right)}{
                \lambda_{\vb{j}(i)}(\vb{r}_i, \nu_i)} .
    } \\
\end{aligned}
\end{equation}
The corresponding weight returned by the backward procedure is
\begin{equation}
    \omega(\vb{S}) =
        \lambda_\text{in}(\vb{r}_I, \nu_I) \prod_{i=1}^n{
            \frac{\sigma_{\vb{j}(i)}(\nu_i)}{\sigma_{\vb{j}(i)}(\nu_{i-1})}
            \omega_{c,\vb{j}(i)}(\nu_{i-1}, \nu_i) .
        }
\end{equation}
By definition, for each collision vertex of index $i$, we have
\begin{equation}
\begin{aligned}
    \frac{\sigma_j(\nu_i)}{\lambda_j(\vb{r}_i, \nu_i)} &={}
        \frac{\sigma_j(\nu_{i-1})}{\lambda_j(\vb{r}_i, \nu_{i-1})}, \\
    p_j^*\left(\nu_{i-1}, \vb{u}_{i-1}; \nu_i, \vb{u}_i\right)
        \omega_{c,j}(\nu_{i-1}, \nu_i) &={}
        p_j\left(\nu_i, \vb{u}_i; \nu_{i-1}, \vb{u}_{i-1}\right) . \\
\end{aligned}
\end{equation}
Thus, it is verified that $p^* \omega = p$, i.e. the weighted backward
trajectories yield back the forward \ac{PDF} as per
equation~\ref{eq:proba-trajectoire}.

Note that algorithm~\ref{al:generation-inverse} generates two additional
trajectory topologies, which are not part of $T$, via the stop condition of
line~\autoref{line:arret-interface-inverse} or
line~\autoref{line:arret-absorption-inverse}. The first case corresponds to the
intersection of an interface, such as the external border $\mathcal{E}$. In this
case, these trajectories could be associated with an external surface source,
such as an external flux of photons, which is not considered here. The second
topology represents backwards-absorbed trajectories. These trajectories do not
have an equivalent in the forward formulation. They result from the effective
treatment of absorption processes, which randomly eliminate some photons to
ensure the correct \ac{PDF} for T-trajectories. If absorption is considered to
be continuous, this branch of the algorithm is not reached. However, this leads
to additional variability in the weights by a factor of $P_s[\Lambda_a] \leq 1$.

\subsection{Generalisation \label{sec:generalisation}}

Let us now generalise the previous algorithms to more complex setup. But, before
delving into the details, let us emphasize that the forward
(\autoref{al:generation-directe}) and backward (\autoref{al:generation-inverse})
transport procedures are almost identical. They both use the same functions for
the transport between vertices (\textproc{CrossSections},
\textproc{DistanceToVertex}, and \textproc{DistanceToInterface}), but with
opposite directions. The main difference between the two algorithms is in the
treatment of inelastic collisions, which call on distinct functions (and
weightings), \textproc{RandomiseCollision} or \textproc{RandomiseConstrained}.
Thus, in practice one can implement a single bi-directional procedure that is
valid in both cases, adding a few conditional branches. However, for clarity we
presented two distinct algorithms in the previous sections.

\subsubsection{Realistic setup}

For a realistic setup, Monte~Carlo trajectories may pass through different media
with distinct compositions before reaching the collection surface. For instance,
the propagation volume may consist of a soil-air binary with an interface
described by topography data. Furthermore, the detector has a composite
structure that includes inert mechanical elements in addition to its gamma-ray
sensitive area.

In the forward case, a method for simulating trajectories within a complex
geometry is to iterate algorithm~\ref{al:generation-directe}. When encountering
an interface (\autoref{line:arret-1}) that is not the collection surface
$\mathcal{C}$, instead of terminating the procedure (\autoref{line:return-1}),
one restarts from the beginning (\autoref{line:cross-sections-1}), after
updating the composition of the propagation medium (on which
\textproc{CrossSections} actually depends). It is important to note that the
photon then straddles two media, which is numerically problematic. Thus, in
practice, one notifies this interface crossing for the next call to function
\textproc{DistanceToInterface} to exclude the interface that was just crossed.

In the backward case, let us point out that
algorithm~\ref{al:generation-inverse} applies transport weights at the level of
collision vertices. Consequently, the previous method is also valid since the
conditions for stopping on and starting from an interface are of {\em flux} type
(see \citet[section~2.6]{Niess2018}).

Another point is that gamma-ray sources have multiple emission lines, rather
than being mono-energetic. In a forward simulation, the initial photon energy is
selected with probabilities $p_k$, according to the respective activities of
emission lines. In a reverse simulation, if the $p_k$ are independent of the
position $\vb{r}$, the source energy can be pre-selected, similar to a forward
simulation. Otherwise, the source energy can be selected based on an a priori
distribution, $p_{b,k}$, and then the trajectory shall be weighted a posteriori
by the ratio
\begin{equation}
    \omega_s = \frac{p_k(\vb{r}_I)}{p_{b,k}} .
\end{equation}

\subsubsection{Mixed Monte~Carlo} \label{sec:mixed-monte-carlo}

It was previously mentioned that the transport procedure can stop and restart at
any interface crossing. This allows for the simulation of a Monte~Carlo
trajectory to be split into two parts, which presents interesting possibilities.
For instance, consider a set of $T_0$ trajectories generated using the backward
procedure described above. If we want to reduce the collection surface to a new
interface, $\mathcal{C}'$, inscribed under the old surface, $\mathcal{C}$, we
can do so by extending the existing $T_0$ trajectories using a forward
procedure. Let us denote $T'_0$ these extended trajectories.

The $T'_0$-trajectories correspond to sources that are external to the volume
included in $\mathcal{C}$. However, it is also likely that the volume between
$\mathcal{C}$ and $\mathcal{C}'$ contains sources, which are called {\em
internal}. To complete the picture, $T'_0$ should be extended with
$T_1$-trajectories, whose sources are internal. Let us denote by $N_0$ ($N_1$)
the number of events simulated with an external (internal) source. Also note
$A_0$ ($A_1$) the total activity of external (internal) sources. The two sets of
trajectories, $T'_0$ and $T_1$, can be combined by assigning a weight of
\begin{equation}
    \omega_A = \frac{A_i}{N_i},
\end{equation}
to each trajectory based on its origin.

Thus, the backward transport procedure reduces a set of sources distributed
over a volume external to $\mathcal{C}$ to a surface flux that enters
$\mathcal{C}$. This flux is represented by the states $S_F$ that terminate the
trajectories $T_0$, weighted according to $A_0 / N_0$. Therefore, it is
advantageous to use two distinct transport algorithms as follows. To begin, a
contour $\mathcal{C}$ is defined to surround the sensitive area of the detector.
Next, external sources to $\mathcal{C}$ are reduced to an incoming flux using a
backward simulation procedure, such as algorithm~\ref{al:generation-inverse}.
Finally, a forward simulation of the detector's response to the incoming flux
and any internal sources is carried out. The second simulation accounts for the
production of secondary particles.

Let us emphasize that the first backward simulation does not require knowledge
of the internal geometry of $\mathcal{C}$. By contrast, the second forward
simulation requires both the internal and external geometry, as a particle may
pass through $\mathcal{C}$ multiple times before detection. Note that during the
backward simulation, it is necessary to exclude re-entrant trajectories to be
consistent with the forward simulation. Therefore, we have imposed that the
simulation of trajectories from external sources stops as soon as $\mathcal{C}$
is intersected.

\section{\Goupil implementation \label{sec:implementation}}

Let us now discuss the software implementation of the Monte~Carlo transport
algorithms (\autoref{sec:algorithms}) within the \Goupil software library.

\subsection{Software architecture \label{sec:architecture-logicielle}}

\Goupil has been implemented in \Rust~\cite{Rust}. In our view, \Rust is an
excellent language for a Monte~Carlo transport engine, enabling the creation of
dependable software without compromising performances. However, \Rust is not yet
widely used in the scientific community due to its relatively recent
introduction. Therefore, we opted for a more traditional user interface,
combining \Python and \C. That is, the internal {\em Rusty} nature of
\Goupil should be transparent to end-users.

Thus, \Goupil is provided as a \Python~3 module, generated from the \Rust source
using \PyO, and available for download from PyPI~\cite{PyPIGoupil}. The project
source is publicly available (under the terms of the LGPL-3.0 license) from
GitHub~\cite{GitHubGoupil}. The module's functionalities are organised into
objects (\Python classes) according to a structure that can be consulted online
from ReadTheDocs~\cite{RTDGoupil}. Time-sensitive routines are vectorised to
take full advantage of \Rust performances. These routines manipulate
\texttt{numpy.ndarray} at the \Python level. For instance, the
\texttt{TransportEngine.transport} method receives a vector of initial states
($S_I$) as input and transforms it, in-place, into a vector of final states
($S_F$).

In addition, \Goupil can be used with an external geometry engine. This
corresponds to the call to the \textproc{DistanceToInterface} function in the
transport algorithms (\autoref{al:generation-directe} and
\autoref{al:generation-inverse}). For this purpose, \Goupil specifies an input
interface in \C, which is also documented on ReadTheDocs~\cite{RTDGoupil}. The
external geometry engine is loaded from \Rust in the form of a dynamic library,
thus avoiding going through \Python during a Monte~Carlo run. In practice, to
implement this approach, a software adaptation layer needs to be developed
between \Goupil and the external geometry engine. For \Geant, a pre-existing
adapter is included with the \Python \Goupil module.

Finally, let us emphasize that \Goupil is interoperable with
\Calzone~\cite{GitHubCalzone}, a \Geant \Python wrapper that was developed in
conjunction with this work in order to facilitate the building of mixed
simulations, as described in section~\ref{sec:mixed-monte-carlo}. Yet, the two
modules can be used fully independently.

\subsection{Collisions\label{sec:physique}}

The relevant interaction processes for \Goupil have been outlined in
section~\ref{sec:particules-secondaires}. Let us recall that, in the absence of
secondary particles, it is only necessary to consider the interactions between
primary photons and matter. Additionally, interactions that do not lead to an
outgoing photon are grouped together as {\em absorption}. Therefore, we will
only be considering three interaction processes: Compton scattering, Rayleigh
scattering and absorption (by the photoelectric effect or the creation of an
$e^+e^-$ pair).

In section~\ref{sec:algorithms}, collision vertices were represented as black
boxes characterised by their statistical properties, via the \ac{PDF} $p_{c,j}$
(or $p^*_{c,j}$, in the backward case). However, the treatment of these
collisions is not trivial. In practice, collisions represent a potential
bottleneck in Monte~Carlo transport requiring dedicated optimisations. In the
case of forward Monte~Carlo, much work was done to have efficient and accurate
implementations of collisions. In particular, let us emphasize the work done for
the \Penelope software (see e.g. \citet{Baro1995,Salvat2015}), on which \Goupil
is based.

For a backward simulation, only the Compton process needs to be reversed.
Rayleigh scattering is of elastic type, and thus self-inverse. Absorption can
also be considered invariant under the conditions of
algorithm~\ref{al:generation-inverse}, as discussed earlier
(\autoref{sec:algorithms}). Moreover, Compton collisions are the most frequent
interaction process at the $\mathcal{O}(1)\,$MeV energies of interest. Thus,
they deserve a more in-depth discussion hereafter, in
section~\ref{sec:diffusion-compton}.

\subsubsection{Target materials}

Before discussing the specificities of each interaction process, let us
highlight some common properties related to the target materials involved in a
collision. First, \Goupil models materials as perfect gases of atoms, which is a
usual assumption for collision processes where energy transfers are much larger
than the inter-atomic binding energies ($\lesssim 10\,$eV).

The atomic elements' properties are predetermined during library initialization,
following the \ac{PDG}. Thus, a \Goupil material is entirely defined by its
atomic composition, expressed in terms of molar or mass fractions. The material
molar mass $M$ and its cross-section $\sigma_j$, for process $j$, are given as
\begin{equation} \label{eq:propriétés-matériau}
    M = \sum_k{x_k M_k}, \quad
    \sigma_j = \sum_k{x_k \sigma_{jk}},
\end{equation}
where the index $k$ runs over the constituent atomic elements of the material,
and the $x_k$ designate the corresponding molar fractions. Note that the
density of the propagation medium, which may vary spatially, is considered by
\Goupil to be a geometric property, and not intrinsic to the target material.

Secondly, it is generally accepted that, where possible, the collisions
parameters should be computed in advance of running a Monte~Carlo simulation.
This leads to the pre-computation of physics tables for each material, which are
interpolated during the simulation instead of performing time-consuming
calculations on the spot. The material cross-section $\sigma_j$ that was defined
above (\autoref{eq:propriétés-matériau}) is a typical example. This
cross-section is dependent on a single variable, the photon energy $\nu$, which
may change during a collision. But, To calculate $\sigma_j$, a lengthy numerical
integration is usually required. Thus, tabulating $\sigma_j$ saves a significant
amount of computing time, at a modest cost in memory.

\Goupil uses physics tables that depend on a maximum of two variables. We
deliberately avoided using tables with more than two variables due to the large
memory overhead that it would imply. For tables with one variable, \Goupil uses
a Piecewise Cubic Interpolation with the method of \citet{Higham1992}, which is
both simple and efficient. For tables with two variables, a Piecewise Bilinear
Interpolation is used. The tables meshing should also be discussed. \Goupil uses
three types of meshes ($x_i$) for a variable of interest $x$: arithmetic ($x_i -
x_{i-1} = \textrm{cst}$), geometric ($x_i / x_{i - 1} = \textrm{cst}'$), or
irregular. In the latter case, a bisection is used to determine an interval for
the point of interest ($x \in [x_i, x_{i-1})$).

For the physics of \Goupil, accurate atomic tables have been previously
published, including detailed cross-sections and form factors. One such
compilation is the EPDL (\citet{Cullen1997}) atomic tables, which are available
in ENDF format from the EPICS~\cite{EPICS} project. Another option is the
XCOM~\cite{XCOM} project, which distributes atomic cross-section tables with a
precision comparable to EPDL, for \Goupil use cases, but with more compact
meshes. By default, \Goupil uses XCOM results for cross-sections and EPDL
results for form factors. The atomic tables are loaded on demand from a local
directory that can be configured by the user. Note that only the absorption and
Rayleigh scattering processes use external atomic tables. In the case of Compton
scattering, \Goupil generates its own material tables, as detailed hereafter
(\autoref{sec:diffusion-compton}).

\subsubsection{Absorption}

Simulating the absorption process is straightforward. One simply stops the
transport by eliminating the photon. However, determining the cross-section of
this process is more involved. The cross-section of photoelectric interactions
is piecewise continuous, with the points of discontinuity reflecting the
underlying electronic structure. The XCOM tables, which are used by default by
\Goupil, account for this by duplicating discontinuity points, in order to
describe the values of the cross-section on either side of the discontinuity.
When combining these atomic data to construct the cross-section of a material,
$\sigma_j$, according to equation \ref{eq:propriétés-matériau}, all these
discontinuities need to be preserved. This is achieved using the procedure
detailed in \ref{sec:merging-procedure}.

\subsubsection{Rayleigh scattering}

Rayleigh scattering corresponds to the interaction of a photon with the static
Coulomb field of a target atom ($\gamma Z \to \gamma Z$), resulting from the
coherent superposition of atomic constituents (nucleus and bound electrons). To
model this process, \Goupil uses an approach similar to \Penelope. The total
cross-section is calculated from accurate atomic tables, using the method of
\ref{sec:merging-procedure}, as for absorption. However, for the simulation of
collisions, Born's~\cite{Born1969} \ac{DCS} is used, such that
\begin{equation}
\begin{aligned}
    p(\cos\theta; \nu) &\propto{}
        \frac{1 + \cos^2\theta}{2} \left|\frac{F(q, Z)}{Z}\right|^2, \\
    q^2 &={} 2 \nu (1 - \cos\theta),
\end{aligned}
\end{equation}
where $\cos\theta = \vb{u}_i \cdot \vb{u}_{i - 1}$ is the angle between the
incident photon and the outgoing photon, and where $F(q, Z)$ is the atomic
form-factor, which depends on the momentum $q = \|\vb{k}_i - \vb{k}_{i-1}\|$
transferred during the collision. Thus, the anomalous scattering term is
neglected when simulating the dynamic of the collision, but it is included in
the total Rayleigh cross-section.

The scattering angle $\theta$ in a Rayleigh collision is sampled using
algorithm~\ref{al:échantillonage-rayleigh}, detailed in
\ref{sec:rayleigh-sampling}. This algorithm differs from \Penelope
implementation by balancing memory over CPU usage, since Rayleigh collisions are
rare in \Goupil use cases.

\subsection{Compton scattering} \label{sec:diffusion-compton}

Compton scattering corresponds to the interaction between an incident photon and
an individual electron in an atom (i.e. $\gamma e^- \to \gamma e^-$).
\citet{Klein1929} first made a precise prediction of the cross-section of this
process, for a free electron, within the framework of quantum electrodynamics.
This result is also applicable, to a good approximation, for the bound electrons
of an atom when the incident photons are sufficiently energetic. However, when
the energy is below $\nu \lesssim 0.1\,$MeV, it is necessary to consider the
electronic structure of the target atom. In the context of a Monte~Carlo
transport engine, performing a detailed structure calculation would not be
affordable. Therefore, \Goupil relies on an effective model detailed below.

\subsubsection{Forward case}

Following \citet{Baro1994} (see also \citet{Ribberfors1982}), the cross-section
for Compton scattering, differential with respect to the scattering angle
$\theta_i$, can be factored as
\begin{equation} \label{eq:compton-angle}
    \frac{d\sigma}{d\cos\theta_i} = \pi r_e^2
        \left(\frac{\nu_c}{\nu_{i-1}}\right)^2
            M(\nu_{i-1}, \nu_c) S(\nu_{i-1}, \cos\theta_i),
\end{equation}
where $M$ is the matrix element of the interaction with a free electron, such
that
\begin{equation}
    M(\nu_{i-1}, \nu_i) =
        \frac{\nu_{i-1}}{\nu_i} + \frac{\nu_i}{\nu_{i-1}} +
        \left(\frac{m_e}{\nu_i} - \frac{m_e}{\nu_{i-1}} - 1\right)^2 - 1,
\end{equation}
and where $S$ is a scattering function, which encodes the electronic structure
of the target atom. In previous equation~\ref{eq:compton-angle},
$\nu_c(\cos\theta_i; \nu_{i-1})$ stands for the energy of the outgoing photon in
a collision with an electron at rest, solution for $\nu_i$ of
\begin{equation} \label{eq:compton-conservation}
    \frac{m_e}{\nu_{i-1}} - \frac{m_e}{\nu_i} + \cos\theta_i - 1 = 0 ,
\end{equation}
which arises from the conservation of energy-momentum.

\citet{Baro1994} proposed a parametric model for the scattering function
$S(\nu_{i-1}, \cos\theta_i)$, which is accurate without requiring detailed
tabulation of $S$. But, unfortunately, equation~\ref{eq:compton-angle} is not
suitable for inversion, because it is differential according to $\cos\theta_i$
instead of $\nu_i$. Thus, \Goupil simplifies equation~\ref{eq:compton-angle}, as
follow. As an approximation, let us substitute $\theta_C(\nu_i; \nu_{i-1})$ for
$\theta$ in equation~\ref{eq:compton-angle}, where $\theta_C$ is the Compton
scattering angle for free electrons, i.e. the solution for $\theta_i$ of
equation~\ref{eq:compton-conservation}. The resulting simplified cross-section,
differential according to the energy $\nu_i$ of the outgoing photon, writes
\begin{equation} \label{eq:compton-forward}
    \frac{d\sigma}{d\nu_i} = \pi r_e^2 \frac{m_e}{\nu_{i-1}^2}
        M(\nu_{i-1}, \nu_i) S_\nu(\nu_{i-1}, \nu_i),
\end{equation}
where $S_\nu(\nu_{i-1}, \nu_i) =
S\left(\nu_{i},\cos\theta_c(\nu_i;\nu_{i-1})\right)$ is set. For the scattering
function of \citet{Baro1994}, we obtain
\begin{equation}
    S_\nu(\nu_{i-1}, \nu_i) = \sum_k{
        f_k \Theta(\nu_{i - 1} - U_k) n_k \circ p_k(\nu_{i-1}, \nu_i)},
\end{equation}
with
\begin{equation}
\begin{aligned}
    n_k(p) &={} \begin{cases}
        z(-p),    & \text{if } p\leq 0 \\
        1 - z(p), & \text{otherwise}
    \end{cases}, &
    z(p) &={} \frac{1}{2}\exp\left(\frac{1}{2}
        -\frac{\left(1 + 2 J_k p\right)^2}{2}\right) , \\
    p_k(\nu_{i-1}, \nu_i) &={} m_e \frac{E_k^2 - \nu_i U_k}
        {\sqrt{2 m_e E_k^2 + \nu_i U_k^2}}, &
    E_k^2 &={} \left(\nu_{i-1} - U_k\right)\left(\nu_{i-1} - \nu_i\right), \\
\end{aligned}
\end{equation}
where the index $k$ runs on the occupied levels of the atom. The parameters
$f_k$ and $U_k$ correspond to the number of occupied levels and the binding
energy, respectively. Parameter $J_k$ reflects the inverse of the average
momentum of electrons within their orbitals. \Goupil utilises the \Penelope
values for these parameters (file \texttt{pdatconf.p14}), which are dependent on
the atomic number $Z$ of the target element.

By construction, \Goupil model reproduces the total cross-section of
\citet{Baro1994} as well as the angular distribution, $d\sigma/d\cos\theta$.
However, the energy $\nu_i$ of the emerging photon is bounded to $\nu_i \in
\left[\nu_\text{min}, \nu_\text{max}\right]$, just as in a collision with an
electron at rest, where
\begin{equation}
        \nu_\text{min}(\nu_{i-1}) = \frac{m_e \nu_{i-1}}
        {m_e + 2 \nu_{i-1}}, \quad
        \nu_\text{max}(\nu_{i-1}) = \nu_{i-1} .
\end{equation}
\Goupil model does not properly account for energy deposition for incident
photons of energy $\nu_{i-1} \lesssim 10\,$keV. But, at these energies, the mean
free path of photons is largely dominated by photoelectric absorption (see e.g.
\autoref{fig:cross-section}).

By default, \Goupil uses a rejection sampling method in order to sample Compton
collisions, as detailed in \ref{sec:forward-compton-sampling}. Alternatively,
the inverse transform method can also be used, at the cost of a significant
increase in memory usage for a moderate CPU gain.

\subsubsection{Backwards case}

In the backward case, by default, \Goupil uses an adjoint Compton model. The
\ac{DCS} of the adjoint process is defined over the support $[\nu^*_\text{min},
\nu^*_\text{max}]$, as
\begin{equation} \label{eq:adjoint-compton}
    \frac{d\sigma^*}{d\nu_{i-1}} = g(\nu_{i-1}, \nu_i) \frac{d\sigma}{d\nu_i},
\end{equation}
where $\sigma$ represents the \ac{DCS} of the forward processes
(\autoref{eq:compton-forward}), and where
\begin{equation}
    \nu^*_\text{min}(\nu_i) = \nu_i, \quad
    \nu^*_\text{max}(\nu_i) = \begin{cases}
        \frac{m_e \nu_i}{m_e - 2 \nu_i}, & \text{if } \nu_i < m_e / 2 \\
        +\infty, & \text{otherwise}
    \end{cases}.
\end{equation}
The function $g$ is a regularisation term that ensures the convergence of the
total adjoint cross-section, $\sigma^*$. The choice $g = \nu_i / \nu_{i-1}$ is
adequate, and in practice it yields satisfactory results. The corresponding
Monte~Carlo weight is
\begin{equation}
    \omega_c = \frac{\nu_{i-1}}{\nu_i}
        \frac{\sigma^*(\nu_i)}{\sigma(\nu_{i-1})}.
\end{equation}
As in the forward case, by default adjoint Compton collisions are simulated
using a rejection method (see \ref{sec:adjoint-compton-sampling}).

Alternatively, since \Goupil Compton model is invertible as per $\nu_i$,
backward Compton collisions can also be simulated using the functional inverse,
as detailed in \ref{sec:inverse-compton-sampling}. However, this approach
implies generating large tables of quantiles. Therefore, it is not used by
default.

Another point to consider is the inclusion of an external energy constraint,
$\nu_I$, during the backward simulation of a Compton collision.
Algorithm~\ref{al:contrainte-ajustée} requires the forward \ac{PDF} $p_c(\nu_i;
\nu_I)$ (\autoref{line:densité-directe}) and the backward \ac{CDF} $P_c^*(\nu_I;
\nu_i)$ (\autoref{line:densité-cumulée-inverse}) when the constraint is reached.
The former $p_c$ is calculated on the spot by \Goupil. In practice, the CPU
impact is acceptable in this case as the constraint is not frequently reached.
However, in the latter case, it is necessary to tabulate $P^*_c$ beforehand. A
1D tabulation per emission line would suffice. But, \Goupil instead pre-computes
a detailed 2D table based on $(\nu_I, \nu_i)$. This allows the user to
seamlessly modify the value of the external energy constraint, i.e. emission
lines.

\subsection{Monte~Carlo geometry \label{sec:géométrie}}

\Goupil represents the Monte~Carlo geometry as a succession of sectors with
uniform atomic composition but variable density. Determining the actual
distances between sectors (\textproc{DistanceToInterface}) is only partially
managed by \Goupil. For stratified geometries that can be represented using one
or more \ac{DEM}, \Goupil disposes of a built-in implementation of the \Turtle
algorithm, which was previously described in \citet{Niess2020}. See e.g. \Goupil
online documentation~\cite{RTDGoupil} for practical examples of usage. For other
use cases, an external geometry engine can be used, as mentioned previously in
section~\ref{sec:architecture-logicielle}.

\Goupil implements two density models. The first model is a uniform density,
while the second model is an exponential density gradient. The latter is
typically used to represent the density within a layer of the atmosphere. This
model is characterised by an axis $\vb{n}_\rho$ and an attenuation length
$\lambda_\rho$. The density $\rho$ at a point $\vb{r}$ is expressed as
\begin{equation} \label{eq:gradient-densité}
    \rho(\vb{r}) = \rho_0 \exp\left(\frac{(\vb{r} -
        \vb{r}_0) \cdot \vb{n}_\rho}{\lambda_\rho}\right) ,
\end{equation}
where $\rho_0$ is the density in a reference plane containing the point
$\vb{r}_0$. The corresponding grammage function writes
\begin{equation}
    X(d;\vb{r},\vb{u}) = \begin{cases}
        \lambda' \left(
            \exp\left(\frac{d}{\lambda'}\right) - 1
        \right) \rho(\vb{r}) & \text{if } \vb{u}\cdot \vb{n}_\rho \neq 0 \\
        d \rho(\vb{r}) & \text{otherwise}
    \end{cases} ,
\end{equation}
where
\begin{equation}
    \frac{1}{\lambda'} = \frac{\vb{u}\cdot\vb{n}_\rho}{\lambda_\rho} .
\end{equation}
This grammage is trivially invertible according to $d$, for fixed $\vb{r}$ and
$\vb{u}$, where the inverse corresponds to the \textproc{DistanceToVertex}
function of the transport algorithms.

\section{Validation and performances \label{sec:validation}}

\Goupil has undergone various validation tests, two of which are presented below
(available from the source, under the {\tt examples/geant4} and {\tt
examples/mixed} folders). The corresponding simulations were conducted on the~\texttt{HPC2}
computing grid located at the Clermont-Auvergne~\cite{Mesocentre} mesocentre.
The grid has a heterogeneous structure, primarily consisting of Intel Xeon
compute nodes running at speeds between 2 and $2.6\,$GHz.

\subsection{Test~1}

The first test involves implementing a simulation in \Geant and exporting the
geometry to \Goupil using its C interface. The results obtained with both
transport engines are then compared.

\subsubsection{Test~1~description}

\begin{figure}[th]
    \center
    \includegraphics[width=\textwidth]{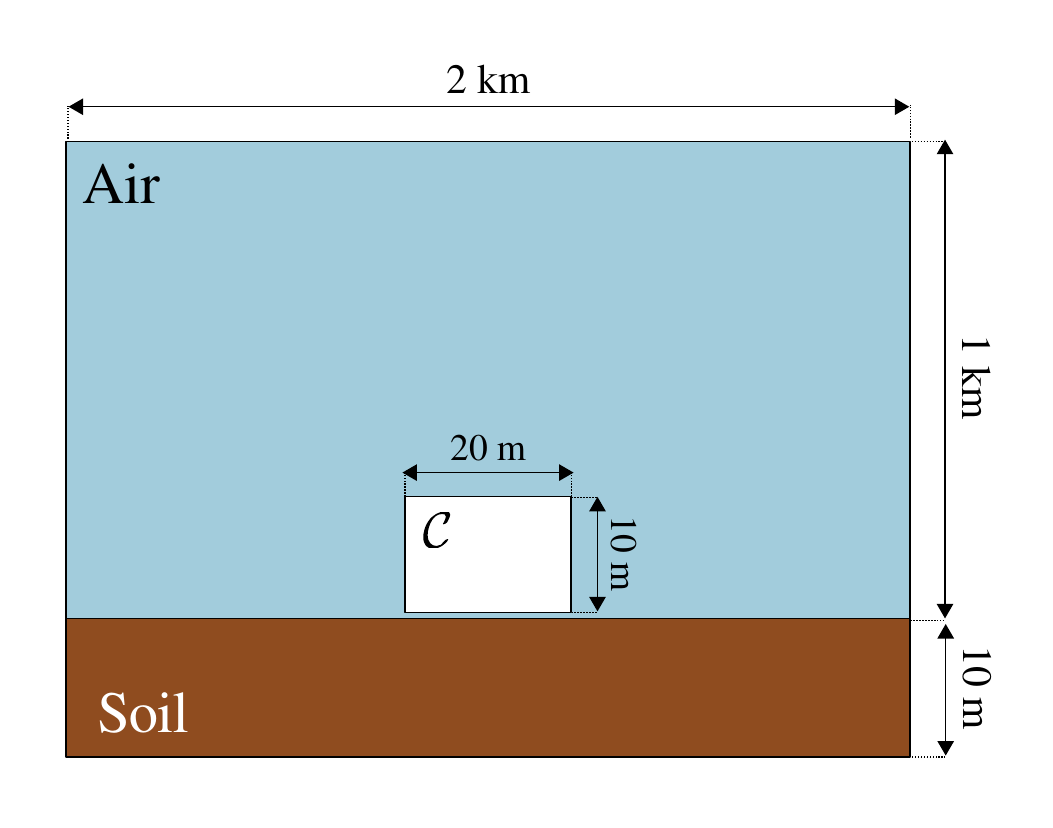}
    \caption{Schematic cross-section of the test geometry in the $(xOz)$ plane.
    The white internal parallelepiped represents the collection volume delimited
    by the surface $\mathcal{C}$. The dimensions of the air, soil, and
    collection volumes are not to scale.
    \label{fig:test-geometry}}
\end{figure}

To compare \Geant and \Goupil, a simple geometry (illustrated in figure
\ref{fig:test-geometry}) is considered. The simulation volume consists of two
boxes superimposed along the $(Oz)$ axis, representing the atmosphere (air) and
the soil (rock). These boxes have a square base of $2\times2\,$km$^2$ along the
$(Ox)$ and $(Oy)$ axes. They are filled with \texttt{G4\_AIR}
($1.205\,\text{mg}/\text{cm}^3$) and \texttt{G4\_CARBONATE\_CALCIUM} (limestone
soil, $2.8\,\text{g}/\text{cm}^3$) respectively (materials from the \Geant-NIST
database).

The detector is modelled as a box also with a square base, measuring
$20\times20\,$m$^2$, and $10\,$m in height. It is placed at the centre of the
simulation volume, $5\,$cm above the soil. The dimensions of the detector in
the simulation are significantly larger than those of a gamma-ray spectrometer,
which is typically only ten centimetres. However, to compare simulation results,
a sufficient number of photons must be collected. This is challenging CPU-wise
with a realistic target size, in a forward simulation. Therefore, we have
artificially increased the extent of the collection volume.

The gamma-ray sources are uniformly distributed throughout the air volume,
excluding the collection volume. Their total activity has been set at $4\cdot
10^{10}\,$Bq (corresponding to a volume activity of $10\,\text{Bq}/\text{m}^3$
of air). The emission lines of these sources are provided in table
\ref{tab:spectrum}. The synthetic experiment consist in counting the photons
collected on the surface $\mathcal{C}$ that surrounds the detector box.

\subsubsection{Comparison~of~performances}

Table~\ref{tab:statistics} provides a summary of the Monte~Carlo data generated
on the HPC2 grid, as well as the respective performances of
the \Geant (version 11.4.1), \Goupil forward and \Goupil backward simulations.
For \Geant, we used Penelope physics (\texttt{G4EMPenelopePhysics}) because it
is similar in principle to \Goupil. However, this choice of physics leads to
simulation times 4.3 times longer than with the standard physics of \Geant
(\texttt{G4EmStandardPhysics}). Note that during the \Geant simulation, we track
all secondary particles to check their contribution to the total flux, as
discussed previously (\autoref{sec:particules-secondaires}). Deactivating the
tracking of these secondary particles makes the \Geant simulation 20 times
faster. Consequently, in the absence of secondary transport and with its
standard physics, \Geant and \Goupil forward have comparable CPU performances.

Considering the time $\Delta t_1$ required for collecting a photon (see
table~\ref{tab:statistics}), \Goupil backward is $\sim$$5 \cdot 10^{4}$ times
faster than \Goupil forward. This impressive gain essentially results from an
increase in the Monte~Carlo efficiency, $\epsilon$. In a forward simulation,
only 3 gamma-rays out of 100\,000 generated reach the collection surface
$\mathcal{C}$ (which is $20\,$m large). During a backward simulation, $39\,\%$
($33\,\%$) trajectories are compatible with a source located in the atmosphere
(soil), $24\,\%$ are backwards-absorbed, and the remaining $4\,\%$ are invalid
trajectories because they turn back towards $\mathcal{C}$.

\begin{table}[th]
    \caption{Statistics of Test~1 Monte~Carlo simulations conducted on the HPC2
    computing grid of the Clermont-Auvergne mesocentre. The table indicates the
    number of generated events $N$, the Monte~Carlo efficiency $\epsilon$
    (i.e. the ratio of the number of gamma-rays collected on $\mathcal{C}$ to
    the number generated), the average simulation time $\Delta t_0$ per
    generated event, and the average time $\Delta t_1 = \Delta t_0 /
    \epsilon$ per collected gamma-ray. Version~11.2.1 of \Geant was used with
    \texttt{G4EMPenelopePhysics}. Note that contrary to \Goupil, \Geant
    simulations include secondary particles.
    \label{tab:statistics}}
\center
\begin{tabular}{lllll}
\toprule
    Simulation & $N$ & $\epsilon$ & $\Delta t_0$ & $\Delta t_1$ \\
\midrule
    \Geant             & $1.5 \cdot 10^{10}$ & $3.0\cdot10^{-5}$ & $1.7\,$ms    &
        $56.6\,$s \\
    \Goupil (forward)  & $1.0 \cdot 10^{11}$ & $3.0\cdot10^{-5}$ & $19.9\,\mu$s &
        $0.67\,$s \\
    \Goupil (backward) & $2.0 \cdot 10^9$    & $3.9\cdot10^{-1}$ & $5.8\,\mu$s &
        $14.9\,\mu$s \\
\bottomrule
\end{tabular}
\end{table}

Let us point out that \Goupil has also been tested with realistic ground
topographies described by \acp{DEM} containing approximately one million nodes,
instead of the simple geometry used for comparison with Geant4. In these cases,
\Goupil maintains excellent performance in backward mode by using a {\tt
StratifiedGeometry} object (\autoref{sec:géométrie}), which implements Turtle's
ray-tracing algorithm (\citet{Niess2020}). \\

\subsubsection{Comparison~of~results}

Let us finally compare the simulations results, particularly concerning the
energy spectrum of photons collected on $\mathcal{C}$. This spectrum is in the
form
\begin{equation}
    \phi(\nu) = B(\nu) + \sum_{k=1}^{n}{R_k \delta(\nu - \nu_k)},
\end{equation}
where $B$ is a piecewise continuous function (over the intervals $[\nu_{k},
\nu_{k+1}]$) representing the background of scattered photons, and where the
right-hand sum runs over the $n=11$ emission lines of energies $\nu_k$,
resulting in photo-peaks of amplitudes $R_k$.

\begin{figure}[th]
    \center
    \includegraphics[width=\textwidth]{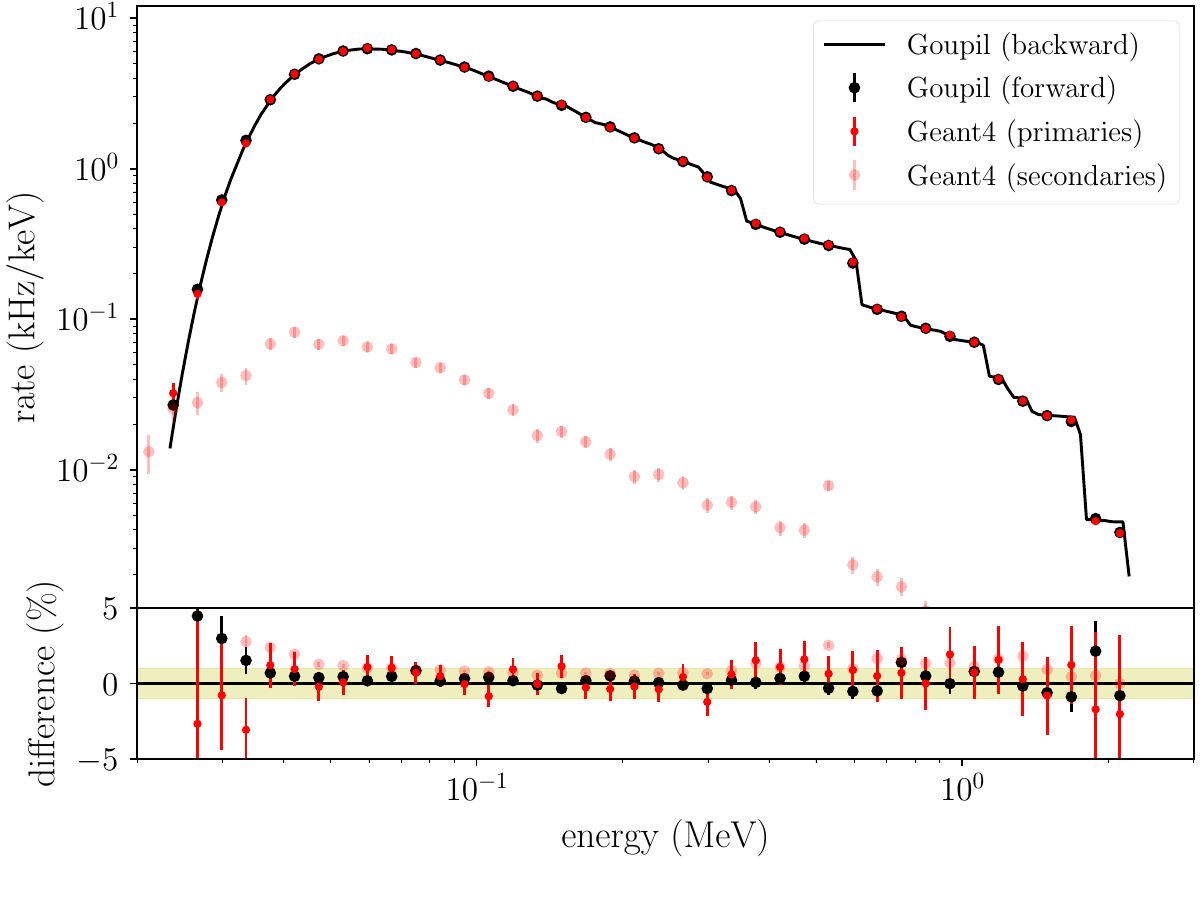}
    \caption{Background energy spectrum ($B$) of gamma-rays collected on the
    inner surface ($\mathcal{C}$) of the Test~1 geometry. The secondary flux
    ($\gamma$, $e^-$, $e^+$) obtained with \Geant is also shown for comparison.
    The error bars indicate Monte~Carlo uncertainties at $68\,\%$ confidence.
    The bottom inset displays the relative deviation w.r.t. the results obtained
    with \Goupil used in backward mode. The yellow band indicates a relative
    deviation of less than $1\,\%$.
    \label{fig:comparison-energy}}
\end{figure}

Figure~\ref{fig:comparison-energy} shows the results obtained for the background
component $B$. The spectra obtained with \Goupil forward or backward are
consistent. The total background obtained in both cases agrees within
$1.5\,\sigma$, with an uncertainty of $0.4\,\%$ on the difference. Similarly,
the results obtained with \Geant are in excellent agreement with \Goupil
($0.4\,\sigma$ with respect to the backward mode) when considering only primary
photons. Secondary particles ($\gamma$, $e^-$, $e^+$) represent less than
$1\,\%$ of the total background flux, which is consistent with the results
presented previously in section~\ref{sec:particules-secondaires}.

\begin{figure}[th]
    \center
    \includegraphics[width=\textwidth]{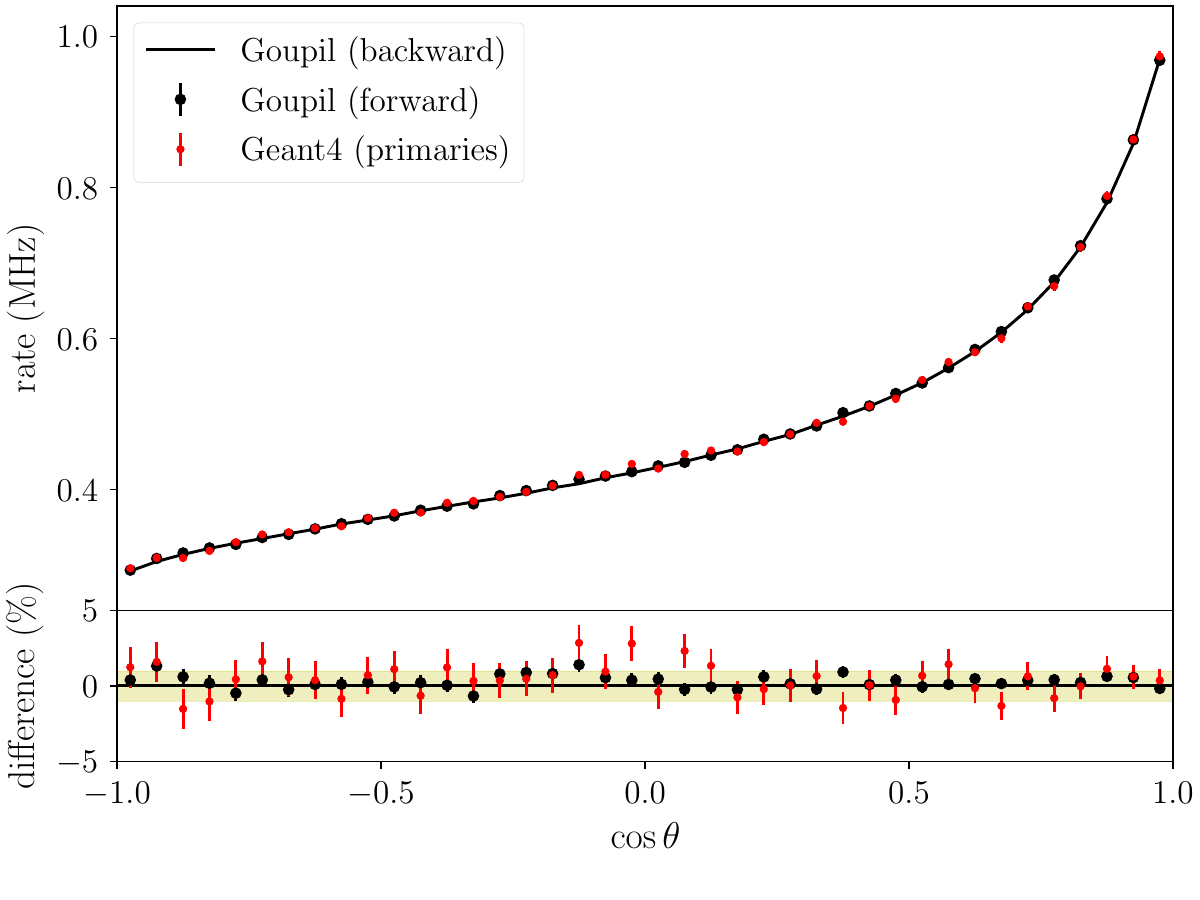}
    \caption{Deflection angle $\theta$ of background photons collected on the
    inner surface ($\mathcal{C}$) of the Test~1 geometry, relative to the
    direction of emission at the source. Monte~Carlo uncertainties are indicated
    by error bars at $68\,\%$ confidence. The lower inset shows the relative
    deviation w.r.t. the results obtained with \Goupil used in backward mode.
    The yellow band indicates a relative deviation of less than $1\,\%$.
    \label{fig:comparison-angle}}
\end{figure}

In addition, figure~\ref{fig:comparison-angle} shows the distribution of the
deflection angle (w.r.t. the emission direction) for the background component.
As for the energy spectrum, no significant differences are observed between the
various simulations.

\begin{figure}[th]
    \center
    \includegraphics[width=\textwidth]{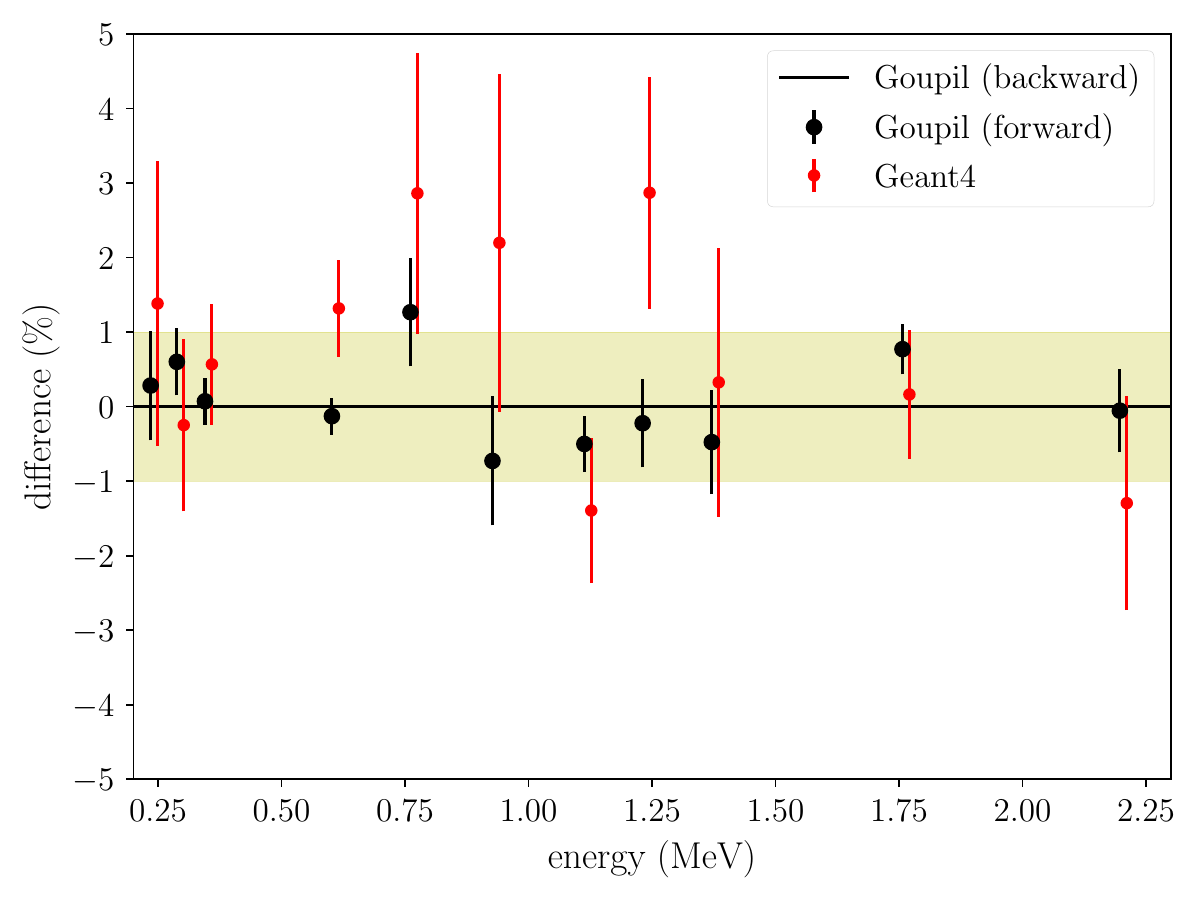}
    \caption{Relative differences (w.r.t. \Goupil backward) in the intensities
    $R_k$ of the photo-peaks observed on the collection surface $\mathcal{C}$ of
    the Test~1 geometry. Error bars indicate Monte~Carlo uncertainties (at
    $68\,\%$ confidence). In the case of \Goupil backward, these uncertainties
    are less than $0.5\,\permil$. The yellow band indicates a relative deviation
    of less than $1\,\%$.
    \label{fig:comparison-discrete}}
\end{figure}

Table~\ref{tab:rate-discrete} lists the values for the photo-peak intensities,
$R_k$, estimated from a weighted average of all simulation results (\Geant,
\Goupil forward and backward). Figure~\ref{fig:comparison-discrete} shows the
relative differences to \Goupil backward for photo-peaks. As for the background
component, forward simulations results (\Geant, \Goupil) are consistent with
backward ones, considering Monte~Carlo uncertainties. Moreover, relative
deviations between \Goupil forward and backward are within $1\,\%$, which aligns
with our accuracy objectives. \\

\begin{table}[th]
    \caption{Photo-peak intensities ($R_k$) in the energy spectrum of gamma-rays
    collected on the inner surface ($\mathcal{C}$) of the Test~1 geometry.
    Reported results are the weighted average of the \Geant, \Goupil forward and
    \Goupil backward simulations. Monte~Carlo uncertainties on the weighted
    average are also indicated.
    \label{tab:rate-discrete}}
    \center
    \begin{tabular}{ll}
\toprule
    Energy, $\nu_k$  & Intensity, $R_k$ \\
    (MeV) & (kHz) \\
\midrule
$0.242$ & $ 7.41$ ($\pm 0.3\,\permil$) \\
$0.295$ & $19.94$ ($\pm 0.2\,\permil$) \\
$0.352$ & $41.03$ ($\pm 0.1\,\permil$) \\
\midrule
$0.609$ & $64.76$ ($\pm 0.1\,\permil$) \\
$0.768$ & $ 7.70$ ($\pm 0.3\,\permil$) \\
$0.934$ & $ 5.31$ ($\pm 0.4\,\permil$) \\
$1.120$ & $27.78$ ($\pm 0.2\,\permil$) \\
$1.238$ & $11.34$ ($\pm 0.3\,\permil$) \\
$1.378$ & $ 8.24$ ($\pm 0.4\,\permil$) \\
$1.764$ & $35.65$ ($\pm 0.2\,\permil$) \\
$2.204$ & $12.77$ ($\pm 0.4\,\permil$) \\
\bottomrule
\end{tabular}

\end{table}

%% --DIFF_COLOR

\subsection{Test~2} \label{sec:test2}

The second test consists in simulating the response of a scintillation-based
gamma-ray detector to radio-isotopes scattered within its surrounding
environment. This test illustrates the mixed simulation scheme discussed in
section~\ref{sec:mixed-monte-carlo}, using \Goupil in backward mode to simulate
the gamma-ray flux in the vicinity of the detector, and then using \Geant to
simulate the detector response to this flux. The results of the mixed simulation
are compared with an end-to-end \Geant forward simulation. \\

\subsubsection{Test~2 description}

The gamma-ray detector is a $7.6\,\mathrm{cm} \times 7.6\,\mathrm{cm}$
cylindrical NaI(Tl) detector, modelled following \citet{DucTam2017} (using the
manufacturer settings, provided by table~1). The detector is immersed directly
in water, with the omission of any readout electronics, power supply, and
waterproof container. While this may not be realistic, these additional elements
are not relevant for the present synthetic test. The detector is immersed in a
dense medium, rather than air, to ensure the end-to-end forward simulation is
CPU-viable on the HPC2 grid.

The source radio-isotopes are distributed uniformly over a $1\,\mathrm{m}$
radius sphere centered on the NaI(Tl) scintillator volume, restricted to the
water volume (i.e. excluding the detector elements). The source emission lines
are taken from \autoref{tab:spectrum} with a unit volume activity of
$\mathcal{A} = 1\,\mathrm{Bq}/\mathrm{m}^3$.

The detector response is assumed to be proportional with a resolution driven by
the photoelectron statistics (see e.g. \citet{Knoll2010}). Thus, the
reconstructed energy, denoted by $\nu_r$, follows a normal distribution, as
\begin{equation}
    \mathcal{N}\left(\nu_r; \Delta\right) = \frac{1}{\sqrt{2 \pi} \sigma}
        e^{-\frac{\left(\nu_r - \Delta\right)^2}{2 \sigma^2}},
\end{equation}
where $\Delta$ is the actual energy deposited in the NaI(Tl) scintillator, as
determined by the Monte Carlo simulation. The resolution factor $\sigma$ can be
expressed as
\begin{equation}
    \sigma(\Delta) = 2\sqrt{2 \ln 2}\, \epsilon_0 \sqrt{\Delta \Delta_0},
\end{equation}
where $\Delta_0 = 0.6617\,\mathrm{MeV}$ is the $\ce{^{137}Cs}$ major emission
line. The energy resolution is set to $\epsilon_0 = 6.7\,\%$, according to a
least square fit of the experimental values reported by \citet{DucTam2017}.

A Monte Carlo estimate of the differential counting rate is given by
\begin{equation} \label{eq:reconstructed-energy}
    \frac{d R}{d \nu_r} \simeq \frac{1}{N} \sum_{i=1}^{N}{
        \omega_i\left(\Delta_i\right) \mathcal{N}\left(\nu_r; \Delta_i\right)},
\end{equation}
where the summation runs over the $N$ generated Monte Carlo events with Monte
Carlo weights $\omega_i$. It should be noted that the previous
\autoref{eq:reconstructed-energy} directly provides the differential counting
rate, w.r.t. $\nu_r$, in the form of a \ac{KDE}. Thus, we do not randomise
the reconstructed energy event by event, which would unnecessarily increase the
Monte Carlo variance and also require a histogram. Instead, we repeat
applying \autoref{eq:reconstructed-energy} by varying $\nu_r$ over a grid of
values (linearly spaced).

The end-to-end forward simulation is carried out using
\Calzone~v1.1.1~\cite{GitHubCalzone} (i.e. \Geant~11.2.1), with Penelope
Physics. In this case, the Monte Carlo weight simply writes as
\begin{equation}
    \omega_i\left(\Delta_i\right) = \begin{cases}
        V_0 \mathcal{A} I  & \text{if } \Delta_i > 0, \\
        0                  & \text{otherwise},
    \end{cases}
\end{equation}
where $V_0$ is the source volume, $\mathcal{A}$ its volume activity and $I =
1.597$ its total gamma-ray intensity (see \autoref{tab:spectrum}). In the mixed
case, intermediary photons are generated over an envelope tightly bounding the
detector. This envelope encompasses the scintillator volume as well as an
aluminium oxide reflector ($2$-$3\,\mathrm{mm}$ thick) and an aluminium housing
($1.5\,\mathrm{mm}$ thick). As outlined in section~\ref{sec:mixed-monte-carlo},
the intermediary photons are then transported backwards using \Goupil, which
provides the Monte Carlo weights $\omega_i$, and forward simulated through the
detector using \Calzone, which provides the energies $\Delta_i$ deposited in the
NaI(Tl) scintillator.

\subsubsection{Comparison~of~performances}

The statistics and performances of the Test~2 Monte~Carlo simulations are
summarised in \autoref{tab:statistics-mixed}, employing the same metrics as for
Test~1. Notably, the mixed simulation is approximately $10^3$ times faster than
the end-to-end forward simulation. While the gain is less pronounced than in an
in-air detector scenario (i.e. Test 1), it remains appreciable. This outcome was
anticipated, given that the source and detector are more comparable in size to
those in Test~1. It is also noteworthy that the \Goupil backward transport stage
accounts for only a few percent of the total CPU time in the mixed case.
Consequently, the relative CPU gain in mixed mode is seldom influenced by the
\Geant physics list.

\begin{table}[th]
    \caption{Statistics of Test~2 Monte~Carlo simulations conducted on the HPC2
    computing grid of the Clermont-Auvergne mesocentre. The table indicates the
    number of generated events $N$, the Monte~Carlo efficiency $\epsilon$ (i.e.
    the ratio of the number of gamma-rays depositing energy in the NaI(Tl)
    scintillator to the number generated), the average simulation time $\Delta
    t_0$ per generated event, and the average time $\Delta t_1 = \Delta t_0 /
    \epsilon$ per detected gamma-ray. Version~1.1.1 of \Calzone was used (i.e.
    \Geant~11.2.1) with Penelope Physics.
    \label{tab:statistics-mixed}}
\center
\begin{tabular}{lllll}
\toprule
    Simulation & $N$ & $\epsilon$ & $\Delta t_0$ & $\Delta t_1$ \\
\midrule
    Forward & $9.5 \cdot 10^{10}$ & $8.9\cdot10^{-4}$ & $0.18\,$ms &
        $0.20\,$s \\
    Mixed   & $2.2 \cdot 10^8$    & $3.0\cdot10^{-1}$ & $59\,\mu$s &
        $0.20\,$ms \\
\bottomrule
\end{tabular}
\end{table}

In mixed mode, $64\,\%$ of the intermediary photons are absorbed before ever
reaching the scintillator, during the forward simulation. Of the remaining
$36\,\%$ of photons, $82\,\%$ successfully sample a source during the backward
stage, resulting in an overall Monte~Carlo efficiency of $30\,\%$.

Substituting a $1\,\mathrm{km}$ radius sphere of air for the $1\,\mathrm{m}$
radius water source has no effect on the overall Monte~Carlo efficiency in mixed
mode. On the contrary, in the case of an end-to-end forward simulation, only
$14$ energy deposits were collected out of $1.2 \cdot 10^{10}$ generated
gamma-rays. This indicates that the CPU-gain would exceed $\sim$$10^8$ for a
scintillator completely immersed in air. Yet, the important point in practice,
is that such simulations are simply unattainable with conventional methods and
computing resources, whereas with the mixed procedure one can already achieve
decent results (thousands of detected events per second) using only a laptop.

\subsubsection{Comparison~of~results}

\begin{figure}[th]
    \center
    \includegraphics[width=\textwidth]{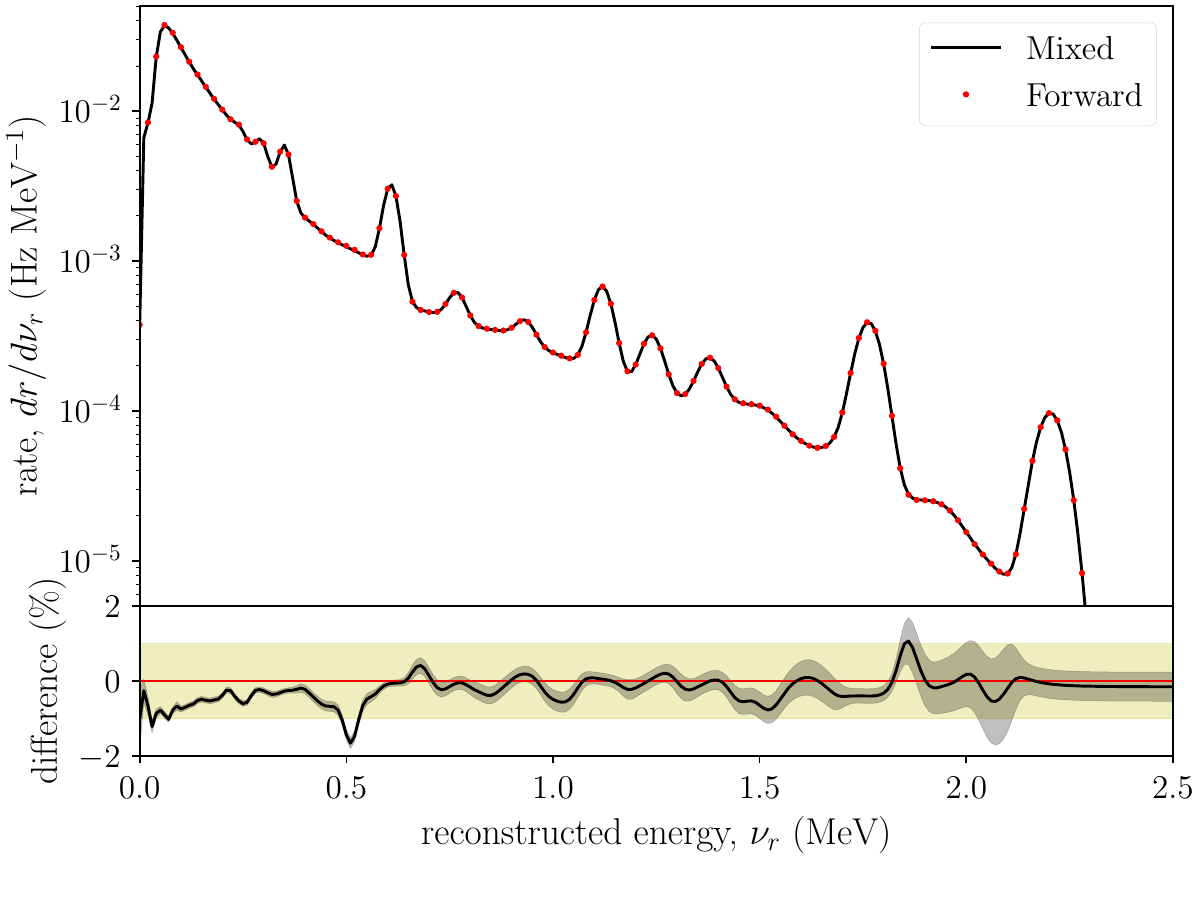}
    \caption{Differential counting rates ($dR/d\nu_r$) obtained with the Test~2
    geometry. The bottom inset displays the relative deviation of the mixed
    procedure w.r.t. the end-to-end forward \Geant simulation. The yellow band
    indicates a relative deviation of less than $1\,\%$. The black shaded area
    represents the Monte~Carlo uncertainties at $68\,\%$ confidence (which are
    correlated due to the convolution with the detector response).
    \label{fig:comparison-test2}}
\end{figure}

The differential counting rates $d R / d \nu_r$ obtained with both procedures
are reported on figure~\ref{fig:comparison-test2}. The mixed procedure
reproduces the end-to-end forward results to within $1\,\%$, with the exception
of the $511\,$keV region, corresponding to gamma-rays emitted by positrons
annihilating in the water volume surrounding the detector (which cannot be
simulated by \Goupil, since secondaries are neglected).

Finally, it is worth noting that the generation of intermediate photons directly
on the scintillator volume (instead of on the detector envelope) leads to an
additional bias of $\sim$$1\,\%$. While the two volumes only differ by a few
millimetres of aluminium, which is sufficient to stop a $1\,$MeV electron, the
secondaries produced in this thin layer (containing the reflector and housing)
do slightly contribute to the detector response.

\section{Conclusion}

In this article, we presented a simple modification
(\autoref{al:contrainte-ajustée}) to our backward Monte~Carlo transport
algorithm (\citet{Niess2018}) allowing for discrete energy sources. This
modification has been implemented in the \Goupil software library for the
Monte~Carlo transport of low-energy gamma photons, typically from radioactive
isotopes. The modified algorithm performs well with comparable computing times
per event ($\Delta t_0$) to the forward mode (see e.g.
table~\ref{tab:statistics}), and an accuracy within one per cent.

\Goupil has a clear advantage over conventional transport engines due to its
high Monte~Carlo efficiency in backward mode. With forward simulations, when the
source volume exceeds the detector size, the efficiency may decrease
significantly. This is particularly evident for gamma-rays transported through
the atmosphere, with Monte Carlo efficiencies, $\epsilon$, that can drop as low
as $\mathcal{O}\left(10^{-9}\right)$. On the contrary, in backward mode \Goupil
maintains efficiencies of $\mathcal{O}(1)$.

It is important to note that \Goupil is not intended to provide a complete
framework, such as a Monte~Carlo simulation of the response of a gamma-ray
spectrometer. Instead, generalist software packages (e.g. \Fluka, \Geant, \MCNP
or \Penelope) might be utilised for this purpose. \Goupil is a specialized
software package that significantly accelerates the transport of gamma photons
in certain circumstances, as previously discussed. It is expected that \Goupil
is used in conjunction with a detector model or another simulation tool. For the
latter case, \Goupil provides a \C interface that allows for the injection of an
external geometry engine. The {\tt goupil} Python module includes an adapter to
\Geant that utilises this \C interface.

In particular, \Goupil is interoperable with \Calzone~\cite{GitHubCalzone}, a
\Geant \Python wrapper that was developed in conjunction with this work in order
to facilitate the building of mixed simulations. An example of such a mixed
simulation, employing a scintillation detector, was presented in
section~\ref{sec:test2}. It was demonstrated that, apart from the $511\,$keV
photopeak resulting from positron annihilation, the mixed simulation procedure
reproduces the outcomes of an end-to-end forward simulation while achieving
events rates of a few kHz (see $\Delta t_1${} in \autoref{tab:statistics-mixed})
with a $\sim$$2.3\,$GHz CPU. However, it should be recalled that, in order to
achieve such performances, the source region must substantially encompass the
detector.

This paper presented a comprehensive academic description of \Goupil, covering
its algorithms and implementation. For practical guidance on using \Goupil,
please consult the available online documentation~\cite{RTDGoupil}.

\section*{CRediT authorship contribution statement}

{\bf Valentin Niess}: Conceptualization, Methodology, Software, Validation,
Formal analysis, Writing -- Original Draft, Writing -- Review \& Editing,
Visualization. {\bf Kinson Vernet}: Software, Validation, Formal analysis,
Writing -- Review \& Editing, Visualization. {\bf Luca Terray}:
Conceptualization, Resources, Writing -- Review \& Editing, Project
administration, Funding acquisition.

\section*{Declaration of Generative AI and AI-assisted technologies in the
writing process}

During the preparation of this work the authors used \DeepL in order to
translate the manuscript from French to English. After using this tool, the
authors reviewed and edited the content as needed and take full responsibility
for the content of the publication.

\section*{Acknowledgements}

We gratefully acknowledge support from the Mésocentre Clermont-Auvergne of the
University of Clermont Auvergne for providing computing resources needed for
this work. This is contribution no. 673 of the ClerVolc program of the
International Research Center for Disaster Sciences and Sustainable Development
of the University of Clermont Auvergne. The analysis of Monte Carlo data has
been done with numpy~\cite{Harris2020}. Validation figures have been produced
with matplotlib~\cite{Hunter2007}.

\appendix
\section{Constrained backward collision \label{sec:proof-constrained}}

In order to prove that algorithm~\ref{al:contrainte-ajustée} is correct, it is
instructive to consider the more general procedure described below
(\autoref{al:roulette-russe}). With the latter method, a constrained backward
collision procedure is constructed from an unconstrained one, noted
$\textproc{RandomiseBackward}$, using a Russian Roulette method. For
simplicity, in algorithm~\ref{al:roulette-russe} we have omitted the index $j$
of the interaction process and the angular coordinate $\vb{u}$, as they are not
relevant (see \cite[Appendix~D]{Niess2018} for example).

\begin{algorithm}[h]
    \caption{
Constrained Backward Collision using a Russian Roulette method.\label{al:roulette-russe}
}
    \begin{algorithmic}[1]
    \Require $\nu_I > \nu_i > 0$, $q \in [0, 1]$,
            $\mathcal{R}$ (a random stream).
        \State $\zeta \gets \Call{Open01}{\mathcal{R}}$
        \If{$\zeta \leq q$} \label{line:branche-1}
            \State $p_c \gets \Call{PdfForward}{\nu_i, \nu_I}$
                \label{line:pdf-direct}
            \State $\omega_c \gets \frac{p_c}{q}$
            \State \Return $(\nu_I, \omega_c)$
        \Else \label{line:branche-2}
            \Loop \label{line:boucle-optionelle}
                \State $(\nu_{i-1}, \omega_c) = \Call{RandomiseBackward}{
                    \nu_i, \mathcal{R}}$
                \If{$\nu_{i-1} < \nu_I$}
                    \State $P^*_c \gets \Call{CdfBackward}{\nu_I, \nu_i}$
                        \label{line:cdf-inverse}
                    \State $\omega_c \gets \frac{\omega_c P^*_c}{(1 - q)}$
                        \label{line:ponderation-contrainte}
                    \State \Return $(\nu_{i-1}, \omega_c)$
                \EndIf
            \EndLoop
        \EndIf
    \end{algorithmic}
\end{algorithm}

Algorithm~\ref{al:roulette-russe} produces two collision topologies, which
depends on the branch of line \autoref{line:branche-1} or
\autoref{line:branche-2}. The first topology corresponds to the collision of a
photon emerging from a point source of energy $\nu_I$. This branch also
determines the end of the backward trajectory simulation, which justifies the
name {\em Russian Roulette}. The second topology simulates the backward
collision of a photon from an energy-distributed source on $[\nu_i, \nu_I)$.
This branch generates an internal collision vertex of the trajectory.
Algorithm~\ref{al:roulette-russe} is correct provided that it assigns the
appropriate weighted probability densities to each branch, following equation
\ref{eq:ponderation-trajectoire}. For the first branch, this condition is
trivially satisfied. For the second branch, it is important to note that the
rejection method used at line \autoref{line:boucle-optionelle} selects a
pre-collision energy $\nu_{i-1}$ based on the \ac{PDF}
\begin{equation}
    p_r^*(\nu_{i-1}; \nu_i, \nu_I) = \frac{p^*_c(\nu_{i-1}; \nu_i)}{
        P^*_c(\nu_I; \nu_i) },
\end{equation}
This justifies the weighting applied at
line~\autoref{line:ponderation-contrainte}.

Algorithm~\ref{al:roulette-russe} is thus compliant, in the sense that it
backwards generates the correct collision topologies and with the correct
weights. However, it requires an adjustment parameter $q$, which affects its
effectiveness. Note that this parameter $q$ can be a function of the
post-collision energy, $\nu_i$, which is actually an input parameter to the
backward procedure. It seems appropriate for $q \to 1$ to let $\nu_i \to
\nu_I$. Furthermore, the energy before collision, $\nu_{i-1}$, might be bounded
by the kinematics of the collision, for a given value of $\nu_i$. Let us denote
this bound $\nu_\text{max}(\nu_i)$. Now, for $\nu_\text{max} < \nu_I$, it is
desirable that $q = 0$, because otherwise the first branch of
algorithm~\ref{al:roulette-russe} could return physically impossible events.
Although these events have zero weight, it is not efficient to generate them.

A suitable option, which meets the aforementioned criteria, is to equal $q$ and
the rejection probability,  as
\begin{equation}
    q(\nu_i, \nu_I) = 1 - P_c^*(\nu_I; \nu_i) .
\end{equation}
This choice eliminates the weighting of
line~\autoref{line:ponderation-contrainte} but most importantly it also let us
eliminate the inner loop (\autoref{line:boucle-optionelle}). This results in
algorithm~\ref{al:contrainte-ajustée}.

\section{Continuous absorption \label{sec:continuous-absorption}}

A valid backward-absorption method shall restore the appropriate weighted
probability density for $T$-trajectories, as per $p^* \omega = p$. That for, it
is worth noticing that the absorption cross-section does not appear in the
collision term $c$ which is used to express $p$ (see
\autoref{eq:proba-trajectoire}). Thus, absorption could be treated as a
continuous process rather than a discrete collision. To clarify, let $\Lambda_c$
be the mean free path obtained by excluding absorption processes, and let
$\vb*{\Lambda}_c$ be the corresponding vector of $\Lambda_j$ values, where $j$
indexes processes. Assuming no absorption, a backward simulation procedure
verifies $p^*[\vb{\Lambda}_c] \omega[\vb{\Lambda}_c] = p[\vb{\Lambda}_c]$ for
any $T$-trajectory. By definition,
\begin{equation}
    P_s[\Lambda] = P_s[\Lambda_a] P_s[\Lambda_c] ,
\end{equation}
where $\Lambda$ is the total mean free path. Since the collision term $c$ is
independent of the absorption processes,
\begin{equation}
    p[\vb*{\Lambda}] = P_s[\Lambda_a] p[\vb*{\Lambda}_c] .
\end{equation}
Therefore, by applying an additional weighting
\begin{equation}
    \omega \gets P_s[\Lambda_a]\, \omega,
\end{equation}
one reproduces the \ac{PDF} $p[\vb{\Lambda}]$.

In practice, the absorption weight, $\omega_a = P_s[\Lambda_a]$, can be built
up step-by-step during the simulation, by weighting the flight between two
successive vertices by the probability of survival of the photon according to
the absorption processes alone. Note that this continuous treatment of
absorption processes can be applied during both backward and forward
simulations (resulting in a non-analogue forward simulation).

\section{Backward photo-peaks \label{sec:photopeaks-proof}}

Backward generated photo-peak trajectories differ from forward ones only by an
additional incomplete collision vertex. Thus, the collision factor of a backward
trajectory writes
\begin{equation}
    c^*(\vb{S}) = c^*_0(S_I) c(\vb{S}),
\end{equation}
where $c^*_0(S_I) = 1 / \lambda_\text{in}(\vb{r}_I, \nu_I)$, since inelastic
collisions are considered as a single process in this case, and since these
collisions are not simulated. Thus,
\begin{equation}
    p^*(\vb{S}) = \frac{p(\vb{S})}{\lambda_\text{in}(\vb{r}_I, \nu_I)},
\end{equation}
which completes the proof.

\section{Merging of atomic cross-sections \label{sec:merging-procedure}}

Let $x_i^k$ denote the sequence of energies, ordered by increasing values, at
which the atomic-cross section $\sigma^k$ was computed (the process index $j$
being omitted). Thus, at a discontinuity one has $x_i^k = x_{i+1}^k$, but
$\sigma_i^k \neq \sigma_{i+1}^k$. The combined mesh $y_\ell$, representing the
sequence of energies for the material cross-section $\sigma$, is constructed as
the increasing ordered sequence of all the $x_i^k$ values, merged over $k$.
Furthermore, let us require that a point $y_\ell$ is duplicated if and only if
it corresponds to a discontinuity in at least one of the $\sigma^k$. Once the
combined mesh has been constructed, the values $\sigma_\ell$ of the material
cross-section at mesh points are computed according to equation
\ref{eq:propriétés-matériau}. This sometimes implies interpolating $\sigma^k$ at
$y_\ell$. In addition, at the left-hand side (right-hand side) of a
discontinuity in $\sigma$, one must take care to use the corresponding left
(right) value of $\sigma^k$.

\section{Sampling of Rayleigh collisions \label{sec:rayleigh-sampling}}

To sample $q^2$ and $\cos\theta$ in a Rayleigh collision, a rejection method is
employed. Let us observe that the function $f = |F / Z|$ is bounded by the
function $g$ defined as
\begin{equation}
    g(q, Z) = \frac{q^2_0(Z)}{q^2_0(Z) + q^2}, \quad \mathrm{where} \quad
    q^2_0(Z) = \sup_{q > 0}{\frac{q^2 f(q, Z)}{1 - f(q, Z)}} .
\end{equation}
The bound $|g|^2$ is an analytical function that allows for easy sampling of
$q^2$ values, such as through the CDF-inverse method. This results in the
algorithm \ref{al:échantillonage-rayleigh} below.

\begin{algorithm}[h]
    \caption{
Sampling of $\cos\theta$ in a Rayleigh collision.\label{al:échantillonage-rayleigh}
}
    \begin{algorithmic}[1]
        \Require $q^2_0 > 0$, $\nu > 0$,
            $\mathcal{R}$ (a random stream).
        \State $x_\text{max} \gets \frac{4 \nu^2}{q_0^2}$
        \Loop
            \State $\zeta \gets \Call{Open01}{\mathcal{R}}$
            \State $x \gets \frac{\zeta x_\text{max}}{
                1 + (1 - \zeta) x_\text{max}}$
            \State $q \gets \sqrt{x q_0^2}$
            \State $f = \Call{FormFactor}{q, Z}$
            \State $g = \frac{1}{1 + x}$
            \State $\cos\theta = 1 - 2 \frac{x}{x_\text{max}}$
            \State $r = \frac{1 + \cos^2\theta}{2} \left[\frac{f}{g}\right]^2$
                \label{line:proba-sélection}
            \State $\xi \gets \Call{Open01}{\mathcal{R}}$
            \If{$\xi \leq r$}
                \State \Return $\cos\theta$
            \EndIf
        \EndLoop
    \end{algorithmic}
\end{algorithm}

This algorithm differs from the implementation of \Penelope, which samples $q^2$
directly according to $|F|^2$. The latter however requires the whole CDF of
$|F|^2$ to be tabulated instead of a single parameter, $q_0^2$. On the other
hand, \Goupil method is slower because the probability of selecting $r$
(\autoref{line:proba-sélection}) is lower by a factor of $f / g \leq 1$. We have
found that this effect is anecdotal for our applications, as Rayleigh collisions
are infrequent compared with other processes. Thus, \Goupil uses
algorithm~\ref{al:échantillonage-rayleigh} because of its small memory footprint
while being sufficiently fast.

\section{Sampling of forward Compton collisions
\label{sec:forward-compton-sampling}}

By default, \Goupil uses a rejection method in order to sample the energy of the
emerging photon, in a Compton collision. Let us point out that the scattering
function $S_\nu \in [0,1]$ decreases with $\nu_i$. Consequently, $\nu_i$ is
sampled according to the matrix element $M$, using the method described in
\citet{Butcher1960}, but with an additional selection factor accounting for
$S_\nu$ as
\begin{equation}
    q(\nu_{i-1}, \nu_i) =
        \frac{S_\nu(\nu_{i-1},\nu_i)}{S_\nu(\nu_{i-1},
        \nu_\text{min}(\nu_{i-1}))} .
\end{equation}
The factor $q$ decreases with $\nu_{i-1}$, making the rejection method
inefficient at low energies. In practice, this is counterbalanced by the fact
that absorption then becomes predominant.

An alternative method for sampling $\nu_i$, which is also implemented in
\Goupil, is to use the inverse transform technique. Numerically, the quantiles
of the distribution function of $\nu_i$ are tabulated for different values of
$\nu_{i-1}$, and then this table is interpolated during the simulation of a
collision. We have found that this method saves a factor of $\sim$2 in
simulation time for transporting a $1/\nu_I$ flux in a uniform medium. However,
in more realistic scenarios, the gain will inevitably be lower due to the
resolution of the geometry impacting performances. Additionally, to achieve
satisfactory numerical accuracy using the inverse transform method, exhaustive
tables containing $\mathcal{O}\left(10^5\right)$ values per material must be
generated. Thus, \Goupil uses the rejection method by default, which is expected
to be sufficiently fast for most use cases.

\section{Sampling of adjoint Compton collisions
\label{sec:adjoint-compton-sampling}}

In the case of the adjoint process (\autoref{eq:adjoint-compton}), the energy
$\nu_{i-1}$ of the incident photon is sampled using a rejection method. The
method is similar to the forward case, but the \ac{PDF} of the generating
process is proportional to $M / \nu_{i-1}^3$ instead of $M$. Additionally, the
selection probability $q$ accounting for the scattering function $S_\nu$ is
modified as
\begin{equation}
    q(\nu_{i-1}, \nu_i) = \frac{S_\nu(\nu_{i-1}, \nu_i)}{
        S_\nu(\nu^*_\text{max}(\nu_i), \nu_i)},
\end{equation}
since $S_\nu$ increases according to $\nu_{i-1}$. The weight, $\omega_c$, is
calculated on the fly from the tabulated forward ($\sigma(\nu_{i-1})$) and
adjoint ($\sigma^*(\nu_i)$) cross-section values.

Let us point out that the inverse transform method can also be used to simulate
the adjoint process with \Goupil, by tabulating the quantiles of the adjoint
\ac{PDF}, $p_c^* = (1 / \sigma^*) d\sigma^* / d\nu_{i-1}$. However, the
rejection method is used by default in order to minimise memory usage, as in the
forward case.

\section{Sampling of backward Compton collisions
\label{sec:inverse-compton-sampling}}

Let $P_c(\nu_i; \nu_{i-1})$ represent the \ac{CDF} of $\nu_i$ in a forward
Compton collision. Assuming that $d\sigma/d\nu_i > 0$, it follows that
$P_c(\nu_i; \nu_{i-1})$ strictly increases with $\nu_i$ on the support
$(\nu_\text{min}, \nu_\text{max})$. It can also be observed that $P_c$ is
strictly decreasing with $\nu_{i-1}$ on the support $(\nu^*_\text{min},
\nu^*_\text{max})$. Thus, the relation
\begin{equation} \label{eq:compton-répartition}
    P_c(\nu_i; \nu_{i-1}) = \zeta, \quad \zeta \in [0, 1],
\end{equation}
establishes a bijection between $\nu_{i-1}$ and $\nu_i$ for a given $\zeta$
value. As a result, the inverse transform method is applicable to both the
forward process (by inverting equation~\ref{eq:compton-répartition} according to
$\nu_i$) and the backward process (by inverting according to $\nu_{i-1}$).
Furthermore, if we denote $\nu^{\shortminus1}(\nu_i; \zeta)$ the inverse
bijection, the Monte~Carlo weight for a backward collision is given by
\citet[lemme~1]{Niess2018} (equation~15).

Technically, the sampling of the backward process is almost identical to the
forward case (\ref{sec:forward-compton-sampling}). However, one needs to
tabulate not only the quantiles with respect to $\nu_{i-1}$, but also the
$\omega_c$ weights. These weights are determined by the slope of the cubic
interpolations of $\nu^{\shortminus1}$ for different $\zeta$ values of the
\ac{CDF}. Like the forward case, these tabulations require a large number of
values to achieve decent numerical accuracy. Therefore, this technique is not
the default method used by \Goupil.

\bibliography{goupil}

\end{document}